\documentclass[twocolumn, twocolappendix]{aastex631}
\usepackage{amsmath}
\usepackage{booktabs}
\usepackage{CJK}
\usepackage[bbgreekl]{mathbbol}
\usepackage{bbold}
\newcommand{\beq}{\begin{equation}}
\newcommand{\baln}{\begin{aligned}}
\newcommand{\eeq}{\end{equation}}
\newcommand{\ealn}{\end{aligned}}
\newcommand{\beqn}{\begin{eqnarray}}
\newcommand{\eeqn}{\end{eqnarray}}
\newcommand*{\vect}[1]{\boldsymbol{#1}}
\newcommand{\Alfven}{Alfv\'{e}n }
\newcommand{\Alfvenic}{Alfv\'{e}nic }

\shortauthors{Li et al.}
\shorttitle{Small-Scale Fields Shape Solar Gamma-Ray Emission}

\begin{document}
\begin{CJK*}{UTF8}{gbsn}

\title{
Small-Scale Magnetic Fields are Critical to Shaping Solar Gamma-Ray Emission
}


\correspondingauthor{Jung-Tsung Li}
\email{li.12638@osu.edu}

\author[0000-0003-1671-3171]{Jung-Tsung~Li (李融宗)}
\affiliation{Center for Cosmology and AstroParticle Physics, The Ohio State University, Columbus, OH 43210, USA}
\affiliation{Department of Physics, The Ohio State University, Columbus, OH 43210, USA}
\affiliation{Department of Astronomy, The Ohio State University, Columbus, OH 43210, USA}

\author[0000-0002-0005-2631]{John~F.~Beacom}
\affiliation{Center for Cosmology and AstroParticle Physics, The Ohio State University, Columbus, OH 43210, USA}
\affiliation{Department of Physics, The Ohio State University, Columbus, OH 43210, USA}
\affiliation{Department of Astronomy, The Ohio State University, Columbus, OH 43210, USA}

\author[0009-0002-1988-0768]{Spencer~Griffith}
\affiliation{Center for Cosmology and AstroParticle Physics, The Ohio State University, Columbus, OH 43210, USA}
\affiliation{Department of Physics, The Ohio State University, Columbus, OH 43210, USA}

\author[0000-0002-8040-6785]{Annika~H.~G.~Peter}
\affiliation{Center for Cosmology and AstroParticle Physics, The Ohio State University, Columbus, OH 43210, USA}
\affiliation{Department of Physics, The Ohio State University, Columbus, OH 43210, USA}
\affiliation{Department of Astronomy, The Ohio State University, Columbus, OH 43210, USA}
\affiliation{School of Natural Sciences, Institute for Advanced Study, Princeton, NJ 08540, USA}

\date{\today}


\begin{abstract}
The Sun is a bright gamma-ray source due to hadronic cosmic-ray interactions with solar gas. While it is known that incoming cosmic rays must generally first be reflected by solar magnetic fields to produce outgoing gamma rays, theoretical models have yet to reproduce the observed spectra. We introduce a simplified model of the solar magnetic fields that captures the main elements relevant to gamma-ray production. These are a flux tube, representing the network elements, and a flux sheet, representing the intergranule sheets. Both the tube and sheet have a horizontal size of order $100~{\rm km}$ and serve as sites where cosmic rays are reflected and gamma rays are produced. While our simplified double-structure model does not capture all the complexities of the solar-surface magnetic fields, such as \Alfven turbulence from wave interactions or magnetic fluctuations from convection motions, it improves on previous models by reasonably producing both the hard spectrum seen by Fermi-LAT at $\text{1--200}~{\rm GeV}$ and the considerably softer spectrum seen by HAWC at near $10^3~{\rm GeV}$. We show that lower-energy ($\lesssim 10~{\rm GeV}$) gamma rays are primarily produced in the network elements and higher-energy ($\gtrsim {\rm few} \times 10~{\rm GeV}$) gamma rays in the intergranule sheets. Notably, the spectrum softening observed by HAWC results from the limited effectiveness of capturing and reflecting $\sim 10^4~{\rm GeV}$ cosmic rays by the finite-sized intergranule sheets. Our study is important for understanding cosmic-ray transport in the solar atmosphere and will lead to insights about small-scale magnetic fields at the photosphere.
\end{abstract}


\section{INTRODUCTION} \label{sec:introduction}

The Sun is a bright and time-steady source of GeV to TeV gamma rays, with the intensity showing a modest anti-correlation with the solar cycle. This emission is proposed to arise from two distinct channels, originating from different locations. The first is emission from the solar halo, arising from electron galactic cosmic-ray (GCR) interactions with solar photons via the inverse Compton process near the Sun. Theoretical predictions \citep{2006ApJ...652L..65M, 2007Ap&SS.309..359O, 2021JCAP...04..004O} reasonably match observations \citep{2011ApJ...734..116A}. The second source of emission arises from the solar disk, where the decay of $\pi^0$ and other mesons produced through nucleon-nucleon collisions occurs as a result of the bombardment of hadronic GCRs on the solar surface. Theoretical predictions for the solar disk emission do not match the observed data. Consequently, in this paper, we focus on understanding this emission.

Observations with the Fermi Large Area Telescope (Fermi-LAT) demonstrate that the solar disk emits gamma rays in the 0.1--200 GeV range with a spectrum, $dN_\gamma/dE_\gamma$, following an $\sim E_\gamma^{-2.2}$ power law \citep{2011ApJ...734..116A, 2016PhRvD..94b3004N, 2018PhRvD..98f3019T, 2022PhRvD.105f3013L}. (For the earlier work with EGRET data, see \cite{2008A&A...480..847O}.) More recently, the High Altitude Water Cherenkov Observatory (HAWC) reported the first detection of gamma-ray fluxes near $10^3$~GeV, finding a $E_\gamma^{-3.6}$ power law \citep{2023PhRvL.131e1201A}. Both instruments reveal an anti-correlation between gamma-ray fluxes and solar activity, with fluxes during the solar minimum being approximately a factor $\sim$ 2 greater than those during the solar maximum. Interestingly, while the entire solar disk emits gamma rays, indicating that the emission is not just a limb effect, the emission distribution reported in \cite{2018PhRvL.121m1103L} shows moderate non-uniformity across the solar surface.

The pioneering theoretical work on this subject is \cite{1991ApJ...382..652S} and \cite{1992NASCP3137..542S}, which focus on the reflection of hadronic GCRs by canopy fields via the magnetic mirroring effect. Their work assumes a simple pressure balance following a $P\left(z\right) \propto B\left(z\right)^2$ scaling relation, where $P\left(z\right)$ and $B\left(z\right)$ are the gas pressure and the magnetic field strength of the flux tube at the height $z$, respectively. They further assume that particle trajectory is governed by the adiabatic invariance, $\cos\theta\left(z\right) = \sqrt{ 1 - B\left(z\right)\sin^2\theta_0 /B_0 }$, where $\theta\left(z\right)$ is particle pitch angle at $z$, $\theta_0$ is the initial pitch angle at the top of the tube, and $B_0$ is the magnetic field strength at the top of the tube. A recent study conducted by \cite{2020MNRAS.491.4852H} suggests a scaling relation of $P \propto B^{3.5\pm 0.1}$ may be more appropriate based on data from the \texttt{Bifrost} magnetohydrodynamics (MHD) simulation code \citep{2011A&A...531A.154G}. Several other approaches and ideas have been suggested in the literature. The studies by \cite{2020PhRvD.101h3011M} and by \cite{2020arXiv200903888L} have integrated the Potential Field Source Surface model of the solar coronal fields into the \texttt{FLUKA} code and implemented numerical simulations of the yields of secondary particles and gamma rays from hadronic GCR showers. \cite{2020APh...11902440G} and \cite{2022ApJ...941...86G} use HAWC data on the cosmic ray shadow of the Sun to deduce the averaged hadronic GCR absorption fraction and the gamma-ray yields. \cite{2023arXiv230517086B} presents a contrasting perspective, arguing that the acoustic-like shock waves in the chromosphere can accelerate protons in the solar gas up to $10^3$~GeV and produce gamma rays that match the HAWC observation. \cite{2017PhRvD..96b3015Z} argues that the solar magnetic fields are negligible for gamma-ray production at high enough energies, in which case the disk emission would be solely from the thin ring of the solar limb; the HAWC data show that this must be at gamma-ray energies above about $10^3$~GeV. Overall, despite extensive theoretical efforts in this domain, no prediction can simultaneously account for the overall gamma-ray flux, the morphology of its emission, and its time variability.

In this paper, we aim to identify the magnetic field structures at the solar surface that are crucial for reflecting hadronic GCRs and to model the gamma-ray emission from these sites. We construct a double-structure solar surface magnetic field model comprising one vertical \emph{flux tube}, representing the magnetic flux tubes that form the network elements, and one vertical \emph{flux sheet}, representing the magnetic flux sheets at the downflow lanes between granules. Both the tube and sheet, which have horizontal sizes of $\sim 100$~km, maintain magnetohydrostatic equilibrium with the surrounding gas. Their field geometries are calculated following the numerical approach in \cite{1986A&A...170..126S}. We inject proton GCRs at the top of the flux tube and sheet, numerically trace their trajectories, and calculate their gamma-ray emission. By doing so, we can determine the fraction of GCRs reflected in the flux tubes of the network elements and the fraction that plunges through to the flux sheets of the intergranule lanes. Our model helps us understand the relative roles of these components in capturing and reflecting GCRs across various energy ranges. Comparison to observations helps probe the solar magnetic environment, even below the surface of the photosphere ($z=0$~km, defined by where the optical depth $\tau_{5000}=1$ for light of wavelength 5000~\AA). In our calculations of the gamma-ray emission, we aim for a precision of a factor of two, which is appropriate given the large dynamic ranges of the variables and the substantial uncertainties.

Last, it is important to note that while solar flares are known to generate gamma rays through reconnection and non-thermal particle acceleration, the emission is generally episodic and does not exceed a few GeV in maximum gamma-ray energy \citep{1987ApJS...63..721M, 1996A&AS..120C.299S, 2014ApJ...787...15A, 2014ApJ...789...20A, 2015ApJ...805L..15P, 2018ApJ...865L...7O, 2018ApJ...869..182S}. Moreover, such events show transient and burst-like behavior and are easily excluded from long-duration gamma-ray observational datasets. As a result, solar flares are not relevant to the time-steady, multi-GeV gamma-ray emission discussed in this paper.

The remainder of this paper is organized as follows. We begin in Section~\ref{sec:overview_field} with an introduction to the solar magnetic field structure. Section~\ref{sec:overall_picture} introduces the double-structure model of gamma-ray emission for the quiet photosphere. In Section~\ref{sec:overall_picture}, we also discuss the potential effect of GCR scattering by magnetic fluctuations. Section~\ref{sec:Magnetostatic_soln} presents the magnetohydrostatic solutions for the flux tube and flux sheet. Section~\ref{sec:simulation} explains our simulation setup for hadronic GCR transport in the double-structure model. Section~\ref{sec:results} presents the simulation results. We conclude in Section~\ref{sec:conclusion}. In Appendix~\ref{sec:absorption_result}, we show the most probable injection polar angles of GCRs for producing the observed gamma rays. In Appendix~\ref{appendix:avg_angle_height}, we show the average gamma-ray emission angle and depth. In Appendix~\ref{appendix:collinearity}, we validate the assumption of collinearity in gamma-ray production.

\section{SIMPLIFIED FRAMEWORK FOR SOLAR SURFACE MAGNETIC FIELDS} \label{sec:overview_field}

In this section, we provide a brief overview of photospheric convection, solar-surface magnetic fields, and open field lines. The reason for this focus will later be explained in Section~\ref{sec:overall_picture}.

\begin{figure*}[t] 
   \centering
   \includegraphics[width=0.99\textwidth]{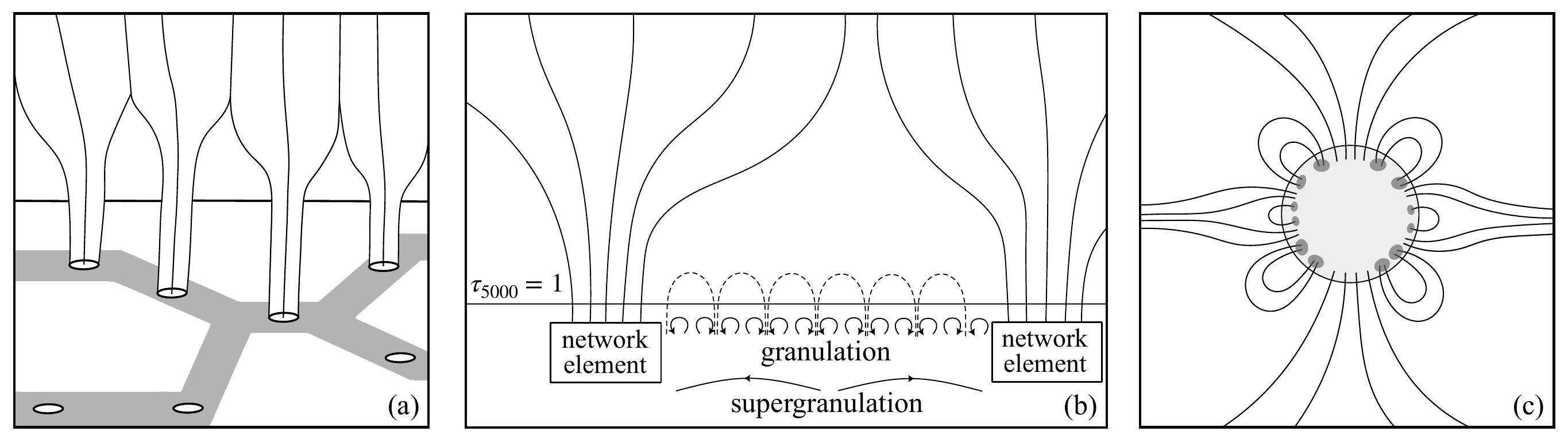}
   \caption{Schematic diagrams (not to scale) of the magnetic network fields and coronal-hole open field lines, with the viewing area progressively expanding from (a) to (c). {\bf(a)} Bird's eye view of solar surface: Vertical magnetic flux tubes arise from granules (white polygons) that reside in the intergranule lanes (gray lanes) along the edges of the supergranule cells. A collection of flux tubes merges at a height of approximately 600--1000~km, forming a network element. {\bf (b)} Vertical 2D slice of the solar surface: The network fields expand laterally and merge with others. The regions between the network elements are filled with granules (small solid arcs with arrows), with magnetic loops (dashed arcs) formed near each granule. {\bf (c)} Solar corona and interplanetary magnetic fields: Lines with one footpoint are the network fields in the coronal-hole regions, forming the open magnetic field lines that extend into interplanetary space. Lines with two footpoints are the closed loops. (a) and (b) are redrawn illustrations based on \cite{2005ApJS..156..265C} and \cite{2009SSRv..144..317W}, respectively.}
   \label{fig:schematic_diagram}
\end{figure*}

Figure~\ref{fig:schematic_diagram} is a schematic diagram of the quiet-photosphere magnetic network structures that form the open field lines. Here we follow the canonical picture outlined in \cite{2005ApJS..156..265C} and \cite{2009SSRv..144..317W}. We begin our discussion with convection in the photosphere, as it is the driving force behind the formation of magnetic flux tubes. Convection near the photospheric surface manifests across various scales. The smallest observable scale is granulation, with a typical horizontal size of about 1000--1500~km (white polygons in Figure~\ref{fig:schematic_diagram}(a)). Each granule cell hosts a convective flow where hot material rises from the cell center, moves horizontally at the top, and descends at the cell edge. Between the granule cells are the intergranule lanes, where the flow is compressed and pushed downward (gray bands in Figure~\ref{fig:schematic_diagram}(a)). At a larger scale is supergranulation, with the typical horizontal size of about 10,000--30,000~km (largest arcs in Figure~\ref{fig:schematic_diagram}(b)). The convective flow responsible for supergranulation occurs on larger scales and at greater depths than the convective flow driving granulation.

Following the variety of scales in convection, the magnetic fields also show a variety of scales. In the case of network fields, the smallest-scale feature is vertical magnetic flux tubes at the photosphere. These tubes, with diameters of approximately 150~km and field strengths of around 1500~Gauss, are situated in the intergranule lanes at the edges of supergranule cells (Figure~\ref{fig:schematic_diagram}(a)). Each flux tube maintains magnetohydrostatic equilibrium with the surrounding gas. As the gas pressure decreases with increasing height, the flux tubes expand horizontally. A collection of nearby flux tubes eventually merges at a height of around 600--1000~km, forming a network element. At the merging height, the horizontal scale of each network element is about 2000--6000~km. Above the merging height, the magnetic fields transition into a thicker flux tube and continue to expand with height (upper-half of Figure~\ref{fig:schematic_diagram}(b)). A second merging occurs when the fields of two adjacent network elements, separated by a horizontal distance of about a supergranule cell, coalesce at the merging height of a few 1000~km above the surface. The magnetic fields continue to extend into the coronal region, becoming parts of the large-scale coronal-hole network fields. These network fields eventually extend into interplanetary space as open magnetic field lines (Figure~\ref{fig:schematic_diagram}(c)), which have their other footprint in the outer structure of the heliosphere.

Magnetic fields between network elements in the quiet photosphere are internetwork fields. Observations indicate that these regions comprise small-scale, mixed-polarity magnetic fields that show loop-like structures a few hundred km above the solar surface (dashed arcs in Figure~\ref{fig:schematic_diagram}(b)) \citep{2007ApJ...670L..61O, 2008ApJ...672.1237L}. These fields are situated above granule cells, with their polarities connecting to the intergranule lanes. Below the photospheric surface, strong vertical magnetic fields are formed in the intergranule lanes as a result of flux expulsion \citep{1963ApJ...138..552P, 1966RSPSA.293..310W, 1981ApJ...243..945G} and field amplification \citep{1978ApJ...221..368P, 1978SoPh...59..249W, 1979SoPh...62...15S} by the granule convective flows. In the near-surface environment, they have sheet-like structures in the intergranule lanes between granules and tube-like structures in the micropores located at the vertices where multiple intergranule lanes converge. Because direct observations below the surface are unavailable, the extent of these structures into the upper convection zone is uncertain. However, valuable insights have been obtained from state-of-the-art numerical simulations of magnetoconvection in the solar photosphere \citep{1996MNRAS.283.1153W, 1998ApJ...499..914S, 1998ApJ...495..468S, 1999ApJ...515L..39C, 2003ApJ...588.1183C, 2005ESASP.596E..65S, 2005A&A...429..335V, 2011A&A...531A.154G, 2012JCoPh.231..919F}. These simulations demonstrate that the coherent structure of flux sheets at the intergranule lanes extends a few hundred km below the surface before being disrupted by turbulent flows.

The discussion thus far restricts to the quiet region of the Sun where magnetic structures near the solar surface are primarily small-scale and driven by granule and supergranule convection cells. Other regions on the solar surface include active regions with more complex magnetic structures. Active regions host strong, large-scale magnetic features, such as sunspots, coronal loops, solar flares, and coronal mass ejections. The activities within the active regions and the solar cycle are thought to stem from the magnetic fields generated by the large-scale solar dynamo emerging from the tachocline at the base of the convection zone.

\section{SETUP OF THE GAMMA-RAY EMISSION MODEL IN THE QUIET PHOTOSPHERE} \label{sec:overall_picture}

In this section, we provide the setup for our model for the time-steady solar gamma-ray emission of the quiet photosphere. Given the extensive range of scales discussed in Section~\ref{sec:overview_field}, simulating GCR propagation and gamma-ray emission in the solar surface environment poses significant challenges. In the following, we explain the approach of our model, which optimizes the prediction of the gamma-ray emission based on appropriate approximations. We review solar modulation and hadronic interactions, then estimate the relevant depths in the photosphere for GCR interactions, outline our key assumptions, define the key magnetic field structures that we model, and discuss the potential effect of the magnetic fluctuations on particle trajectories.

\subsection{Solar Modulation} \label{subsec:solar_modulation}

As the GCR flux has not yet been measured near the Sun, it must be calculated based on the flux near Earth. When hadronic GCRs in interplanetary space propagate toward the Sun, their fluxes decrease due to the interactions with magnetic turbulence in the solar wind, a phenomenon known as solar modulation~\citep{1965P&SS...13....9P, 1968ApJ...154.1011G}. In our previous work in \cite{2022ApJ...937...27L}, we utilized recent magnetic power spectral density measurements from the Parker Solar Probe to calculate the parallel diffusion coefficients. We then showed that the proton GCR flux is reduced by $\lesssim 15\%$ for proton kinetic energy, $E_p^{\rm k}$, in the range of $0.1~{\rm GeV} < E_p^{\rm k} < 1~{\rm GeV}$ when transported from a heliocentric distance of 1~au to 0.1~au. For the higher energy range of GCRs considered here, the effects are even smaller, $\mathcal{O}\left(\lesssim 1\%\right)$ for $E_p^{\rm k} \gtrsim 10~{\rm GeV}$.

The GCR transport from 0.1~au to the solar surface ($\approx 0.005$~au) remains uncertain. Accurate evaluation of this transport requires theoretical models of magnetic turbulence, solar wind acceleration, and coronal loops. This region was not considered in our previous work in \cite{2022ApJ...937...27L} and is beyond the scope of the present work. (See e.g., \cite{2023ApJ...943...21P} for modeling of electron GCR propagation down to the solar surface.) However, we expect that the modulation from 0.1~au to the solar surface should be no more than a few percent for $E_p^{\rm k} \gtrsim 10~{\rm GeV}$ based on the model extrapolation in \cite{2022ApJ...937...27L}. Given that the focus of gamma rays in this work is above 1~GeV, hence proton GCRs above 10~GeV (see below), the modulation effects on our predictions should be no more than $\simeq 10\%$, and are hence neglected.

In this work, we use the GCR spectra measured at 1~au from the Sun, which takes into account the significant modulation effects experienced as GCRs propagate from the heliopause to 1~au, for GCR injection into the solar surface magnetic fields. We have further assumed an isotropic GCR distribution. We use the proton GCR measurements from AMS-02 \citep{2015PhRvL.114q1103A} for $0.5~{\rm GeV} \leq E_p^{\rm k} \leq 1.46\times 10^3~{\rm GeV}$ and from CREAM \citep{2022ApJ...940..107C} for $2.10\times 10^3~{\rm GeV} \leq E_p^{\rm k} \leq 5.33\times 10^5~{\rm GeV}$, linearly interpolating the fluxes between the data points. In Section~\ref{subsec:enhancement_factor}, we describe how we take into account the moderate effect of helium in the GCRs and the Sun.

\subsection{Hadronic GCR Interactions} \label{subsec:collision_type}

The interactions of hadronic GCRs in the solar atmosphere can be understood based on well-established results on how these interact in Earth's atmosphere \citep{2016crpp.book.....G}. Because the GCR energies considered are so large, the details of the atomic states (e.g., ionization) are wholly irrelevant, and even the details of the nuclear states (e.g., hydrogen vs.\ nitrogen) are largely so.

The dominant loss process for proton GCRs is inelastic nucleon-nucleon collisions (``{$pp$ interactions}'' below) that produce pions and other mesons. In our calculations, we use the results of \cite{2006PhRvD..74c4018K}, plus validations with the \texttt{FLUKA} code \citep{2015_Battistoni_fluka, 2022FrP.....9..705A}. Though multiple mesons may be produced, the basic physics is well summarized by $p + p \rightarrow p + p + \pi^0$, followed by $\pi^0 \rightarrow \gamma + \gamma$. The total inelastic hadronic cross section for proton-proton interactions is $\sigma_{\rm pp} \simeq 3 \times 10^{-26}~{\rm cm^2}$, with weak energy dependence for $E_p^{\rm k} \gtrsim 1$~GeV \citep{2022PTEP.2022h3C01W}. The proton mean free path, expressed as a mass column density, is thus $\simeq 55$ g cm$^{-2}$ (about half a meter of water equivalent), meaning that gamma rays are only produced when the GCRs encounter substantial material. The differential cross section can be roughly approximated by assuming that the most important gamma ray has $E_\gamma \sim 0.1 \, E_p^{\rm k}$. Last, these $pp$ interactions may break nuclei, either directly or through induced showers, and this may lead to nuclear de-excitation gamma rays; as those are below about 10 MeV, we ignore them here.

The second-most important --- but ignorable --- loss process is continuous losses due to GCR collisions with electrons. For neutral material, the loss rate due to atomic excitation and ionization is approximately ${\rm 2~MeV/\left(g~cm^{-2}\right)}$ for all materials except hydrogen and ${\rm 4~MeV/\left(g~cm^{-2}\right)}$ for hydrogen, the difference is due to the greater number of electrons per unit mass \citep{2022PTEP.2022h3C01W}. The solar photosphere is largely neutral, but we note that even the losses in fully ionized material (so-called Coulomb losses) are only a few times larger \citep{1998ApJ...509..212S}. These loss rates have only weak dependence on energy in our range of interest. The continuous energy loss of a proton over one mean free path for inelastic nucleon-nucleon interactions is thus $\simeq 0.2$~GeV. Due to our focus on gamma rays above 1~GeV, and hence protons above about 10~GeV, this continuous energy loss is negligible.

\subsection{Relevant Photospheric Depths} \label{subsec:estimate}

Using the proton mean free path, we estimate the possible depth range for interactions, which informs which magnetic field structures are most relevant. For simplicity, here we only assume vertical trajectories, ignoring helical motion. Interactions typically occur when $ \int \sigma_{pp} n_{\rm H}\left(z\right) dz \gtrsim 1$, where $n_{\rm H}\left(z\right)$ represents the hydrogen number density of the gas at height $z$.

Using the mass density $\rho$ of the Sun, which will later be given in Figure~\ref{fig:HSRASP_model}, we estimate that incoming proton GCRs typically penetrate no deeper than $800$~km below the surface. Because the mass density varies exponentially with depth, this is a firm bound. As a result, proton GCR absorption and gamma-ray emission primarily occur within the \emph{photosphere} and \emph{uppermost convection zone}.

In Appendix~\ref{appendix:avg_angle_height}, we show that a full calculation with our model reveals that emission primarily occurs within a height ranging from $-100~{\rm km}$ to $500~{\rm km}$. This is shallower than noted in our estimate because our 3D numerical simulation of particle trajectory takes into account full helical motion, hence a longer particle trajectory at around the reflection height compared to the case for the vertical trajectories. Moreover, to produce outgoing gamma rays, proton GCRs must first be reflected from incoming to outgoing.

\subsection{Key Assumptions of our Model} \label{subsec:fields_of_interest}

The Sun's magnetic environment is complex, with distinctive features ranging from the smallest scales to the size of the solar system. However, when considering how incoming GCRs produce outgoing gamma rays, there are basic facts that allow significant simplifications. First, hadronic GCRs only produce gamma rays when the accumulated mass column density they traverse is very large. Because the matter density in the solar surface varies exponentially, only a narrow range of heights is relevant, as discussed in Section~\ref{subsec:estimate}. This suggests that, as a primary effect, we should consider magnetic fields from network elements and intergranule lanes within this specific height range. Second, there must be magnetic fields strong enough to reflect the very energetic GCRs from incoming to outgoing so that the gamma rays can escape over the whole surface of the Sun, as observed. In comparison, we can reasonably neglect deflections from other field structures, though they may matter at second order through how they change the angular distribution of incoming GCRs.

In our model, we make the following simplifications:
\begin{enumerate}
\item That the solar surface is spherically symmetric, and thus the model presented in this work applies to every small patch on the surface of the Sun.
\item That all GCRs reach the flux tube of the network elements in the quiet photosphere by propagating along the open field lines and network fields.
\item That higher-energy GCRs that penetrate the flux tubes of the network element plunge onto internetwork regions, which consist of vertical flux sheets.
\end{enumerate}
The first simplification allows us to focus on the main parts of the emission process; however, it means we will not be considering the different latitudes of the coronal holes and open field lines. The second simplification aids in understanding how emission occurs in the quiet photosphere, even though we will not explore the relationship between gamma-ray flux, active regions, and solar activity in this paper. The third simplification helps us separate the trajectories of low- and high-energy GCRs and gives us a method to evaluate the fraction of high-energy GCRs entering the internetwork regions.

\subsection{Details of the Double-Structure Model} \label{subsec:model_schematic}

\begin{figure}[t] 
   \centering
   \includegraphics[width=0.98\columnwidth]{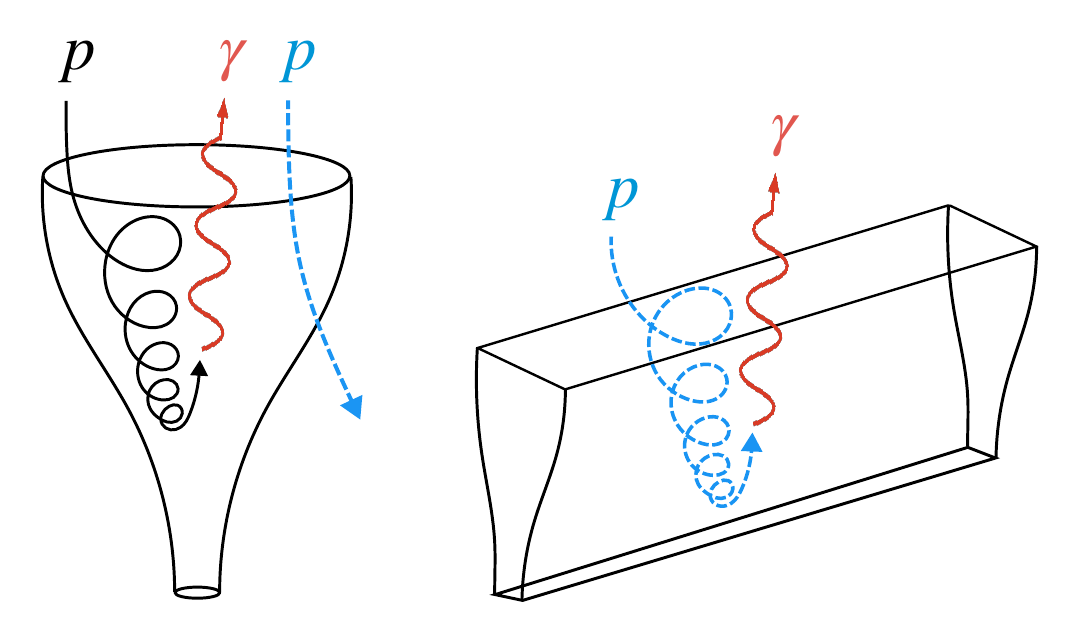}
   \caption{Schematics of the double-structure model. \textbf{Left:} Lower-energy hadronic GCRs (black helix) are magnetically mirrored inside the flux tube, producing outgoing gamma rays (red wavy line). Higher-energy hadronic GCRs (blue curve) plunge through the flux tube with a slight deflection. \textbf{Right:} Higher-energy hadronic GCRs (blue helix) then enter the flux sheet, with a fraction being magnetically reflected, producing outgoing gamma rays.}
   \label{fig:model_schematics}
\end{figure}

Figure~\ref{fig:model_schematics} shows the schematics of the double-structure magnetic fields we use to model GCR propagations. It consists of one isolated flux tube and one isolated flux sheet. The tube represents those forming the network elements, as depicted in Figure~\ref{fig:schematic_diagram}(a). The sheet represents those formed in the intergranule lanes, as depicted in Figure~\ref{fig:schematic_diagram}(b). In our model, this sheet is symmetric along the long horizontal axis and extends indefinitely. The magnetic structures of the flux tube and flux sheet are essential to determining the 3D trajectories of the GCRs. In Section~\ref{sec:Magnetostatic_soln}, we demonstrate the numerical solutions for these structures under the magnetostatic equilibrium with the surrounding gas.

Driven by our assumption that hadronic GCRs propagate towards the Sun along open field lines and network fields, eventually reaching the network elements, we start the GCR trajectories at the top of the flux tube. Hadronic GCRs with an isotropic distribution are uniformly injected across the cross-section surface of the tube at the merging height, set at $z=1600$~km, as shown in Section~\ref{sec:Magnetostatic_soln}. Lower-energy GCRs --- those with Larmor radii, $r_L$, much smaller than field scale height, $\left(\nabla \ln \lvert \vect{B} \rvert \right)^{-1}$ --- are tightly confined by the field lines within the flux tube. As they spiral downward, the magnetic field strength increases due to compression from the surrounding gas, causing the particle pitch angle to approach $90^\circ$. Eventually, the radial component of the field imparts a ``kick'' to the particles, initiating the upward spiral, the process known as the magnetic bottle effect.

On the other hand, higher-energy GCRs --- those with $r_L \gg \left(\nabla \ln \lvert \vect{B} \rvert \right)^{-1}$ --- do not spiral inside the tube; instead, they plunge through the tube with a slight deflection. After they exit the tube, we maintain their polar angles (relative to the vertical axis) while isotropizing their azimuthal angles. These GCRs are then injected into the flux sheet, onto a horizontal surface at $z=600$~km, as depicted in Figure~\ref{fig:schematic_diagram}(b) and Figure~\ref{fig:model_schematics}. Let $L$ be the separation between the two adjacent granule lanes, and $W_{\rm sh}$ be the width of the flux sheet at $z=600$~km. Then the ratio $W_{\rm sh}/L$ is the fraction of those higher-energy GCRs exiting the tube that should be injected into the sheet. For the GCRs that enter the sheet, we calculate their particle trajectories and gamma-ray yields.

At the very highest energies, GCRs are not captured by solar-surface magnetic fields. They will plunge straight through the flux sheets and interact with the Sun but do not produce outgoing gamma rays.

\subsection{Magnetic Fluctuations in the Solar Atmosphere} \label{subsec:Bfluctuation}

The magnetic field structure described in Section~\ref{subsec:model_schematic} is a magnetostatic configuration, which does not contain any magnetic fluctuations. In this subsection, we discuss the potential effect of magnetic fluctuations on particle trajectories.

We distinguish the impacts of magnetic fluctuations originating from three distinct regions within the solar atmosphere in the open field scenario. The first region is the solar corona, where \Alfvenic turbulence occurs. This turbulence is thought to contribute to the heating of the plasma in the coronal region and the acceleration of the solar wind \citep{1987SoPh..109..149T, 1988JGR....93....7T, 1999ApJ...523L..93M, 2007ApJS..171..520C, 2010ApJ...708L.116V, 2011ApJ...736....3V, 2016ApJ...821..106V}. The energy source of the turbulence comes from the \Alfven waves launched from the photospheric surface due to the buffeting of the footprints by solar granulation flows \citep{1998ApJ...509..435V, 2003ApJ...587..458N, 2005ApJS..156..265C}. As these waves propagate outward from the photospheric surface, they interact with the inward-propagating waves arising from wave reflection in the lower coronal region \citep{1980JGR....85.1311H, 1989PhRvL..63.1807V, 1993A&A...270..304V, 1999ApJ...523L..93M, 2002ApJ...575..571D, 2003ApJ...597.1097D, 2011ApJ...736....3V, 2016ApJ...821..106V, 2023arXiv230810389M}. The energy cascade then happens due to the nonlinear interactions between the counter-propagating \Alfven waves, producing microscopic \Alfvenic turbulence at smaller scales (i.e., higher wavenumbers) \citep{1964SvA.....7..566I, 1967PhFl...10.1417K}. The turbulence generated this way in the solar corona is transported away from the Sun by the solar wind into the heliosphere. We point out that the origin of the solar modulation discussed in Section~\ref{subsec:solar_modulation} is due to GCR interaction with this kind of turbulence from the solar wind. The reduction of GCR intensity due to solar modulation within 1~au is small \citep{2022ApJ...937...27L} and thus neglected in this work. Because the gamma-ray flux produced in the corona is much smaller than that produced near the photospheric surface, as discussed in Section~\ref{subsec:estimate}, the magnetic fluctuations in this region are not the primary concern of this work.

The second region is the photosphere and the lower chromosphere, where the gamma rays are most likely to be produced, as shown in Section~\ref{subsec:estimate}. Specifically, we refer to the region from $z=0$~km to $z=1600$~km, above which the density is not high enough to produce significant gamma rays. The magnetic fluctuations in this region are macroscopic MHD transverse waves (fluctuations) driven by photospheric motions and the waves reflected from the lower coronal region. (Specifically, the waves below the merging heights of the flux tubes at the height of 600--1000~km are classified as kink waves, while those above the merging height are identified as \Alfven waves. Both types are referenced as MHD transverse waves in our discussion.) \cite{2005ApJS..156..265C} highlights that the energy cascade efficiency is low in this region in the open field scenario, which is also supported by the findings in \cite{2007ApJS..171..520C}. This low efficiency is evident in Figure~14 of \cite{2005ApJS..156..265C}, which illustrates that the \Alfven wave reflection time, $t_{\rm ref}$, is shorter than the nonlinear outer-scale eddy cascade time, $t_{\rm eddy}$. Following the hierarchy of the energy cascade by \cite{2003ApJ...597.1097D}, a weak energy cascade occurs when $t_{\rm ref} < t_{\rm eddy}$, as \Alfven waves propagate away before the turbulence has sufficient time to develop and heat the plasma. In contrast, the strong energy cascade occurs when $t_{\rm eddy} < t_{\rm ref}$, as there is sufficient time for the \Alfven turbulence to develop. Therefore, the condition in \cite{2005ApJS..156..265C} of $t_{\rm ref} < t_{\rm eddy}$ in the open field regions of the photosphere and chromosphere suggests a weak magnetic power spectrum of the \Alfven turbulence for scales smaller than the outer scale of the driving eddies, i.e., an inefficient process to produce smaller-scale sideway displacements on the magnetic field lines.

This weak turbulence suggests that particle scattering with \Alfven turbulence is likely also weak, though whether it is negligible in the study of solar gamma-ray emission requires a careful evaluation against pitch-angle scattering and magnetic mirroring times. Similar analysis has been conducted in \cite{2001ApJ...549..402M} and \cite{2018ApJ...868L..28E} on the topics of thermal conduction within stochastic magnetic fields and their application to energetic particle transport in solar flares. The two references indicate that if the escape time in the magnetic mirroring region is less than the pitch-angle scattering time, then magnetic mirroring due to the converging fields has a stronger effect than the scattering of the particles on the \Alfven turbulence, and the opposite is true if the scattering time is less than the escape time. However, a precise quantitative analysis of particle pitch-angle scattering requires a detailed magnetic power spectrum of \Alfven turbulence as functions of wavenumber and height in the photosphere and chromosphere. Such information has not been shown in the existing literature. The derivation of this magnetic power spectrum would entail modeling macroscopic MHD and acoustic waves originating from photospheric motions, their reflection at the coronal base and lower corona, nonlinear wave interactions for direct and inverse energy cascade processes, and a plasma heating model that aligns with observational data. These requirements exceed the scope of the present study. As such, our current flux tube and sheet model does not consider microscopic \Alfven turbulence, though this aspect remains a focus for future investigations.

The third region is the upper-most convection zone, situated within a few hundred~km below the photospheric surface. The macroscopic fluctuations of the magnetic fields and the fluids are caused by the convective nature of the granule cells. Even in this convective environment, a coherent structure in tubes and sheets can emerge due to flux expulsion \citep{1963ApJ...138..552P, 1966RSPSA.293..310W, 1981ApJ...243..945G} and field amplification \citep{1978ApJ...221..368P, 1978SoPh...59..249W, 1979SoPh...62...15S}, and it is this coherent structure that we investigate in this study. In contrast, the macroscopic fluctuating parts of the fields cannot be easily obtained through theoretical approaches. They require 2D or 3D magnetoconvection simulations, which exceed the current paper's scope but will be pursued rigorously in future research. In this study, we omit the macroscopic fluctuating component and the microscopic \Alfven turbulence of the magnetic fields, focusing solely on the coherent component of the field configuration in the uppermost convection zone, photosphere, and chromosphere. It is noteworthy that our results, as presented herein, reasonably agree with the observational gamma-ray data, signifying that the double-structure approach in a magnetostatic configuration setting captures key characteristics of the hadronic GCR reflection and the resulting solar gamma-ray emission.

\section{MAGNETOHYDRODYNAMIC SOLUTIONS OF FLUX TUBE AND FLUX SHEET} \label{sec:Magnetostatic_soln}

In this section, we describe the magnetic flux tube and flux sheet structures employed in our model. We first present the equations governing the magnetohydrostatic equilibrium with the surrounding gas, then develop the numerical solutions, for which we follow the approach in \cite{1986A&A...170..126S}. We emphasize that the coordinate variable names used in this section should not be conflated with those used in Section~\ref{sec:simulation} for particle trajectories.

\subsection{Basic Equations} \label{subsec:basic_equations}

\emph{Flux tube:} The flux tube representing the network elements is approximated as a vertical, axisymmetric, untwisted magnetic flux tube that is in magnetohydrostatic equilibrium with the surrounding gas. We describe the flux tube in the cylindrical coordinates ($r$, $\phi$, and $z$). The magnetic field $\vect{B}$ of this geometry has azimuthal symmetry, i.e., $\partial \vect{B}/\partial{\phi} = 0$. By introducing the stream function $\Psi \equiv r A_\phi$ where $A_\phi$ is the $\phi$ component of the vector potential, the field components can be described by 
\beq
    B_r = -\frac{1}{r}\frac{\partial \Psi}{\partial z}, \quad B_z = \frac{1}{r}\frac{\partial \Psi}{\partial r}, \quad B_\phi = 0.
\eeq
The stream function $\Psi$ satisfies the quasi-linear (Grad-Shafranov) equation,
\beq
    \frac{\partial^2 \Psi}{\partial r^2} - \frac{1}{r}\frac{\partial \Psi}{\partial r} + \frac{\partial^2 \Psi}{\partial z^2} = -4 \pi r J,
    \label{eq:NE_tube}
\eeq
where $J$ is the $\phi$-directional current density (with the dimension of [${\rm current/length^2}$]) at the surface between the flux tube and the surrounding gas.

\emph{Flux sheet:} The flux sheet representing the intergranule sheets is approximated as a vertical, untwisted magnetic flux sheet in a magnetohydrostatic equilibrium with the surrounding gas. We describe the flux sheet in Cartesian coordinates ($x$, $y$, and $z$). The magnetic field of this geometry has translational symmetry, i.e., $\partial \vect{B}/\partial x = 0$. The stream function is the $y$ component of the vector potential, i.e., $\Psi=A_y$. The field components are described by 
\beq
    B_y = -\frac{\partial \Psi}{\partial z}, \quad B_z = \frac{\partial \Psi}{\partial y}, \quad B_x = 0,
\eeq
where $\Psi$ satisfies the quasi-linear equation,
\beq
    \frac{\partial^2 \Psi}{\partial y^2} + \frac{\partial^2 \Psi}{\partial z^2} = -4 \pi J,
    \label{eq:GS_sheet}
\eeq
with $J$ the $x$-directional current density at the surface between the flux sheet and the surrounding gas.

\emph{Common features:} 
The magnetic fields at the surfaces of the flux tube and flux sheet are discontinuous and thus form current sheets with surface current density $J^\star$ (with the dimension of [${\rm current/length}$]). In this paper, we refer to this current sheet as the ``flux boundary''. For magnetohydrostatic equilibrium, the total pressure, $P + B^2/8\pi$, must be continuous across the flux boundary, i.e., 
\beq
    P_{i} + B_{i}^2 / 8\pi = P_{e} + B_{e}^2 / 8\pi ,
    \label{eq:P_continuum}
\eeq
where the subscripts $i$ and $e$ refer to the internal and external boundary surfaces of the flux tube or flux sheet. Using Ampere's Law and the pressure continuity in Equation~\eqref{eq:P_continuum}, $J^\star$ can be expressed as
\beq
    J^\star = \frac{2\left(P_{e} - P_{i}\right)}{B_{e} + B_{i}}.
\eeq

To solve the quasi-linear equations in Equations~\eqref{eq:NE_tube} and \eqref{eq:GS_sheet}, we make several simplifications about the surrounding gas. First, we consider an idealized case where the gas temperature $T$ at any given height $z$ is in equilibrium in the horizontal direction, i.e., $T_{i}\left(z\right) = T_{e}\left(z\right) = T\left(z\right)$. Thus, the interior and exterior gas pressure scale heights are equal and are given by
\beq
    H\left(z\right) = \frac{R_\odot^2 k_b T}{G \mu M_\odot},
    \label{eq:scale_height}
\eeq
where $R_\odot$ is the solar radius, $k_b$ is the Boltzmann constant, $G$ is  Newton's constant, $\mu$ is the mean molecular weight, and $M_\odot$ is the solar mass. The interior and exterior gas pressures are then expressed as
\beq
    P_{i}\left(z\right) = P_{i}\left(z_b\right) \exp\Bigg[{-\int_{z_{\rm b}}^z\frac{dz^\prime}{H\left(z^\prime\right)} dz^\prime}\Bigg],
\eeq
\beq
    P_{e}\left(z\right) = P_{e}\left(z_b\right) \exp\Bigg[{-\int_{z_{\rm b}}^z\frac{dz^\prime}{H\left(z^\prime\right)} dz^\prime}\Bigg].
\eeq
where $P_{i}\left(z_b\right)$ and $P_{e}\left(z_b\right)$ are the internal and external gas pressures immediately adjacent to the flux boundary, and $z_{b}$ is the base of the computation box.

Additionally, we assume that at $z=z_{b}$, the vertical magnetic field components inside the flux tube or flux sheet have a uniform distribution, while outside the flux tube or flux sheet, the magnetic fields are zero. As a result, the difference between the interior and exterior gas pressures at the base is expressed as
\beq
    P_{e}\left(z_b\right) - P_{i}\left(z_b\right) = \big[ B_{z}\left(z_b\right)^2 + B_{\perp}\left(z_b\right)^2 \big] / 8 \pi,
\eeq
where $B_{z}\left(z_b\right)$ and $B_{\perp}\left(z_b\right)$ are the vertical and perpendicular $\vect{B}$ components just inside the flux boundary at the base. For the case of the flux tube, $B_{\perp} = B_r$; for the case of the flux sheet, $B_{\perp} = B_y$.

\subsection{Numerical Solutions}\label{subsec:numerical_soln}

We follow the iterative procedure from \cite{1986A&A...170..126S} to solve for $\Psi$ from the quasilinear equations in Equations~\eqref{eq:NE_tube} and \eqref{eq:GS_sheet}. Because of the symmetries of the flux tube and flux sheet, we only solve half of the tube domain (the $rz$ plane) in Equation~\eqref{eq:NE_tube} and half of the sheet domain (the $yz$ plane with $y \geq 0$) in Equation~\eqref{eq:GS_sheet}. In both cases, the boundaries at the top of the computational domains follow the Neumann condition where $\partial \Psi/\partial z = 0$ assuming $B_r=0$. To obtain stable solutions, we have used the \emph{implicit underrelaxation} method introduced in \cite{1980wdch.book.....P}.

\begin{figure}[t] 
   \centering
   \includegraphics[width=0.99\columnwidth]{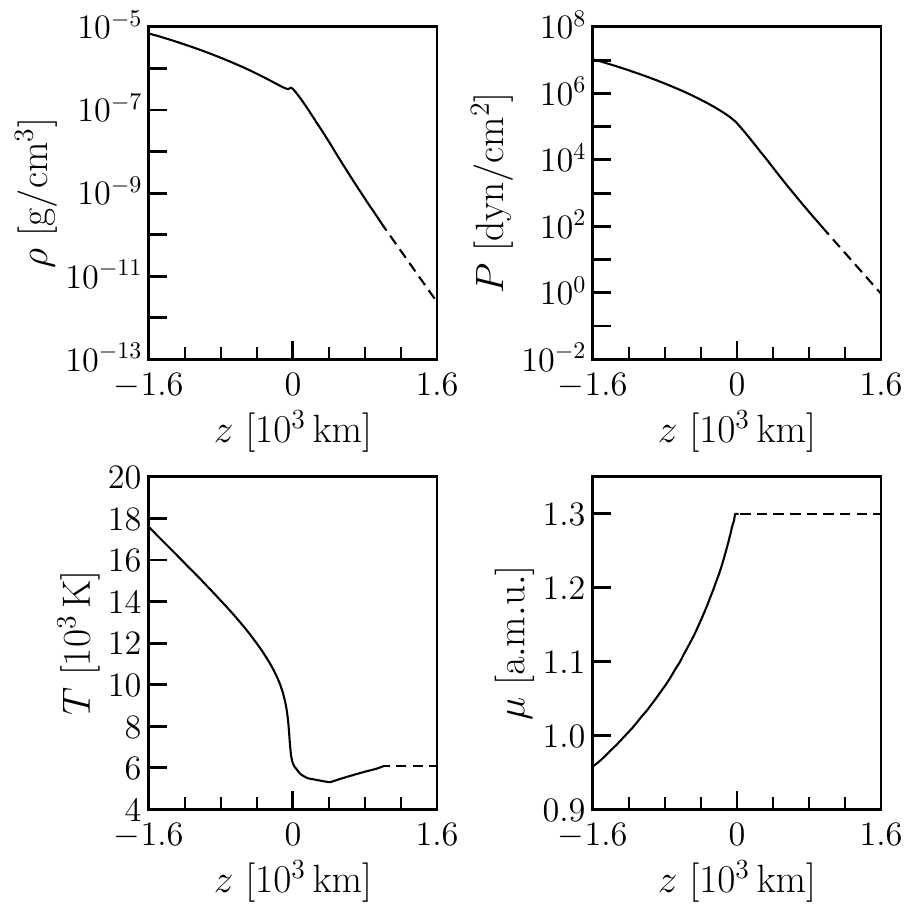}
   \caption{Adopted model conditions for the upper convection zone, photosphere, and chromosphere. Solid lines are data from the HSRASP model. Dashed lines are our extrapolation.}
   \label{fig:HSRASP_model}
\end{figure}

To evaluate the pressure scale height in Equation~\eqref{eq:scale_height}, we must also specify $T$ and $\mu$ as a function of $z$. For this purpose, we use the HSRASP model from \cite{1979ApJ...232..923C}. This model combines the HSRA model from \cite{1971SoPh...18..347G}, describing the gas properties in the chromosphere and photosphere, with the upper convection zone model from \cite{1974SoPh...34..277S}.

Figure~\ref{fig:HSRASP_model} shows the HSRASP model for the upper convection zone, photosphere, and chromosphere. The photosphere layer is located in $0~{\rm km} < z < {\rm few} \times 100~{\rm km}$, with the chromosphere located above and the convection zone below. Because the HSRASP model does not provide $\rho$, $P$, and $T$ above $z=1000~{\rm km}$ and $\mu$ above $z=0~{\rm km}$, we make assumptions to extrapolate these data. First, we set $T = T\left(1000~{\rm km}\right) = 6070~{\rm K}$ for $1000~{\rm km} \leq z \leq 1600~{\rm km}$. This temperature aligns with the model VAL-A of the semi-empirical 1D hydrostatic calculation from \cite{1981ApJS...45..635V}, which shows the quiet-Sun $T\approx 6000$~K in the same range of $z$. Next, we set $\mu = \mu\left(1000~{\rm km}\right) = 1.3$ for $0~{\rm km} \leq z \leq 2000~{\rm km}$. For the region $1000~{\rm km} \leq z \leq 1600~{\rm km}$, we calculate $\rho$ and $P$ under the assumption $\rho\left(z\right) = \rho\left(1000~{\rm km}\right)\exp\left({-\Delta z/H_0}\right)$ and $P\left(z\right) = P \left(1000~{\rm km}\right)\exp\left({-\Delta z/H_0}\right)$ where $\Delta z = z - 1000~{\rm km}$ and $H_0 \equiv H\left(1000~{\rm km}\right)$.

Last, we assume that Sun's atmosphere is comprised of only hydrogen (H) and helium (He) atoms. Heavier elements are ignored since they only account for $\lesssim 2\%$ of the total mass. As a result, the mass fraction of the hydrogen atom is given as
\beq
    X_{\rm H}\left(z\right) = \frac{\mu\left(z\right)^{-1} - A_{\rm He}^{-1}}{A_{\rm H}^{-1} - A_{\rm He}^{-1}},
\eeq
where $A_{\rm H} = 1$ and $A_{\rm He}=4$ are the mass numbers of hydrogen and helium, respectively.

\subsubsection{Flux Tube Solution} \label{subsubsec:tube_soln}

\begin{figure}[t] 
   \centering
   \includegraphics[width=0.23\textwidth]{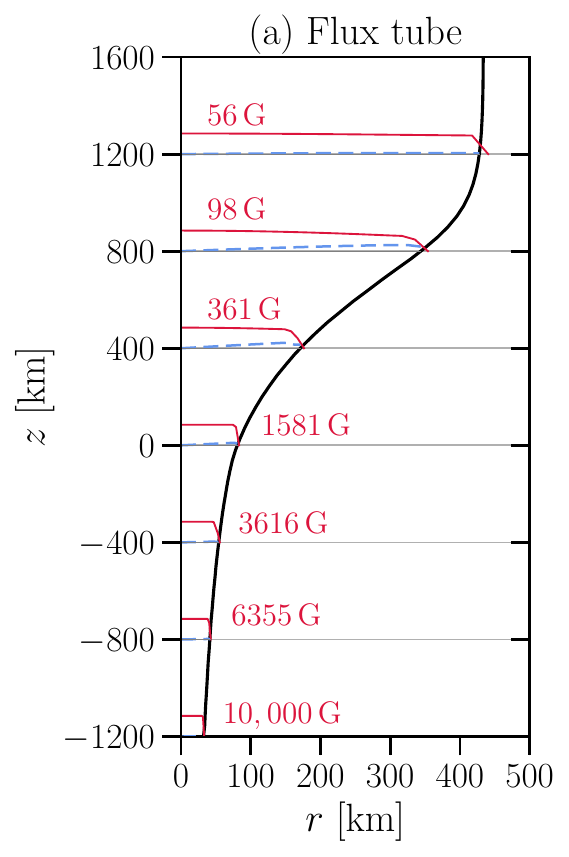}
   \includegraphics[width=0.23\textwidth]{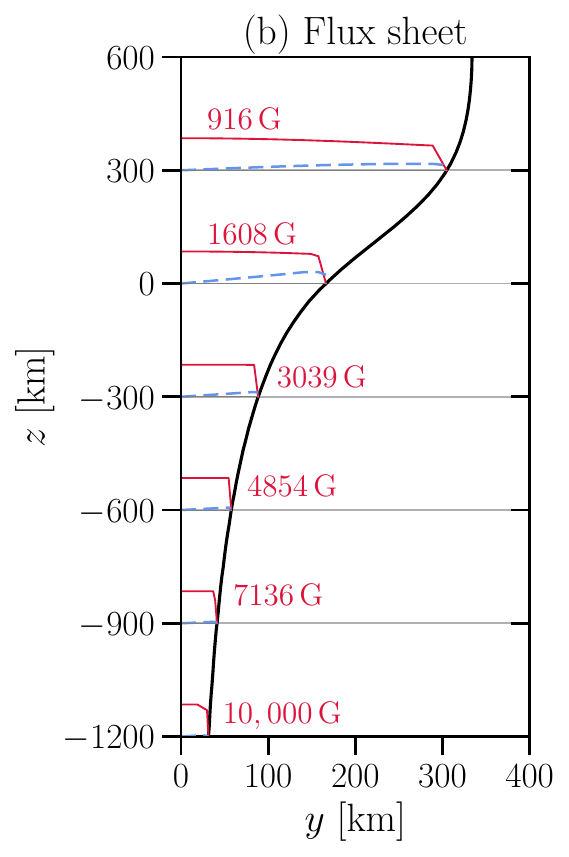}
   \caption{\textbf{(a)}~For the flux tube case: vertical cross-section plane and magnetic field strength variations. The black line denotes the edge of the flux tube. At each height (labeled by a horizontal gray line), the $r$-direction variations of $B_z$ and $B_r$, normalized to the value of $B_z$ at the axis (with their numerical values labeled in red), are shown in the red solid and blue dashed lines, respectively. \textbf{(b)}~For the flux sheet case. (Note the vertical scale in this representation has been compressed by factors of $2.8$ and $2.25$ in relation to the horizontal scales depicted in (a) and (b), respectively. The actual dimensions of the flux tube and sheet are more slender than the depicted structures here.)}
   \label{fig:flux_tube_sheet}
\end{figure}

For the numerical solution of the flux tube, the computational domain is a $2800~{\rm km} \times 500~{\rm km}$ box that is discretized on a rectangular mesh of $65 \times 65$ mesh points. The base of the box is set at $z_{b}=-1200~{\rm km}$. We select an initial radius for the flux tube, $R^\star$, and vertical magnetic field strength, $B_z\left({z_b}\right)$ at the base of the computational domain, with values of $36$~km and $10,000$~Gauss, respectively. This choice of initial conditions gives the radius of the tube at $z=0~{\rm km}$ as 80~km, the filling factor as $3.6\%$, and the axial field strength at $r=z=0~{\rm km}$ as $1580~{\rm Gauss}$. The numerical value of the tube radius at $z=0$~km is close to the photometric measurements reported in \cite{1983SoPh...87..243M}, which shows that the typical flux tube diameter at $z=0$~km for the facular points forming the network fields in the quiet photosphere is 150~km. The numerical values of the filling factor and the axial field strength agree with the network properties revealed by the Stokes $I$ and $V$ lines of Fe-{\footnotesize{I}} reported in \cite{1984A&A...140..185S}, which show that the vertical field component of the network field ranges from 1400~G to 1700~G, and the filling factor is between $3\%$ to $4\%$.

Figure~\ref{fig:flux_tube_sheet}(a) shows the numerical solution for the flux tube, which is invariant in the azimuthal direction. The radius of the tube at the top of the simulation box ($z=1600$~km) is $R_{\rm top} = 430$~km. The figure is divided into groups of subfigures to show the variations for $B_z$ and $B_r$ at different heights. At each height, we find that as $r$ increases, $B_r$ increases while $B_z$ decreases.

Lastly, we emphasize that the magnetic field structures discussed here are located in the photosphere and chromosphere. They are finite-sized and the building blocks of the large-scale coronal and interplanetary magnetic fields. These flux structures located at the photosphere have much stronger field strength ($>1000$~G) than those in the coronal region, which have a field strength of $\sim {\rm few}$~G at $1~R_\odot$ from the photospheric surface~\citep{1987SoPh..109...91P}. The large field strength of the structures we consider, plus their location in dense regions, is essential to redirecting GCRs from inward to outward below the column density of material needed to produce gamma rays.

\subsubsection{Flux Sheet Solution} \label{subsubsec:sheet_soln}

For the numerical solution of the flux sheet, the computational domain is an $1800~{\rm km} \times 500~{\rm km}$ box that is discretized on a rectangular mesh of $65 \times 65$ mesh points. The base of the computational box is set at $z_{b}=-1200~{\rm km}$, as for the tube. We choose the initial half-width of the sheet and the vertical magnetic field strength $B_z\left({z_b}\right)$ at the base of the computation box to be $30$~km and $10,000$~Gauss, respectively. This choice of the initial conditions gives the width of the sheet at $z=0~{\rm km}$ as 330~km, the filling factor as $55\%$, and the vertical field strength at $r=z=0~{\rm km}$ as $1608~{\rm Gauss}$. The half-width of the sheet at the top of the simulation box, $z=600$~km, is given as $y_{\rm top} = 330$~km. Consequently, $W_{\rm sh} = 2 y_{\rm top} = 660$~km. The numerical values of the widths at $z=0$~km and at $600$~km, as well as the vertical field strength at $z=0$~km, are similar to the results obtained from the 3D magnetoconvection simulation presented in \cite{2012JCoPh.231..919F}.

Figure~\ref{fig:flux_tube_sheet}(b) is the numerical solution for the flux sheet, which is invariant along the $x$ direction. (The sheet structure is redundant in the $x$ direction.) The interpretation is the same as Figure~\ref{fig:flux_tube_sheet}(a), except for the blue dashed line denoting $B_y$ variation in this model.

\section{SIMULATION APPROACH FOR GCR PROPAGATION AND GAMMA-RAY EMISSION} \label{sec:simulation}

In this section, we explain our numerical simulation of solar disk gamma-ray emission. Our simulation approach is factorized into four separate stages, taking advantage of the fact that we need to calculate only the effects on the average GCR proton and gamma-ray distributions, as opposed to needing to follow the details of every particle. First, we simulate proton GCR trajectories in the flux tube and flux sheet, recording the accumulated optical depths of these trajectories, but ignoring interactions. Next, we calculate the energy spectrum of gamma rays resulting from $pp$ interactions, based on the proton optical depths. We then integrate the gamma-ray fluxes along the trajectories of the proton GCRs to obtain the total emission flux from the solar disk due to $pp$ interactions. Finally, we incorporate helium contributions to the GCR flux and the solar gas density by using the nuclear enhancement factor.

\subsection{Proton GCR Propagation} \label{subsec:GCR_propagation}

Proton GCR motions follow the Lorentz force equation, 
\beq
    \frac{d\vect{v}}{dt} = \frac{q}{\Gamma\left(\vect{v}\right) m_p \, c} \; \vect{v} \times \vect{B}\left(\vect{r}\right),
\eeq
where $\vect{v}$ is proton velocity, $\Gamma$ is the Lorentz factor, $q$ is proton charge, $m_p$ is proton rest mass, $c$ is the speed of light, and $\vect{B}\left(\vect{r}\right)$ is the local magnetic field at location $\vect{r} = \left(x,y,z\right)$. We do not include the electric field $\vect{E}$ in the Lorentz force equation as its magnitude is negligible compared to the $\vect{v} \times \vect{B} / c$ term. We motivate this within an ideal magnetohydrodynamics framework where $\vect{E} = -\vect{U}\times \vect{B} / c$, with $\vect{U}$ denoting the plasma flow velocity. The typical convective flow speed in the granule cell is $\lvert\vect{U}\rvert \sim 1~{\rm km/s}$, five orders of magnitude smaller than the hadronic GCR speed, $\lvert \vect{v} \rvert \approx c$.

\subsubsection{Injection into the Flux Tube} \label{subsubsec:injection_tube}
We uniformly inject protons across the horizontal cross-section surface of the flux tube at $z=1600$~km. Due to the azimuthal symmetry of the tube, we only inject protons along one of the horizontal axes. We begin the injection process at an axial distance of $r_0=20$~km and progressively increase $r_0$ in increments of $40$~km until it reaches the boundary of the flux tube.

At each $r_0$, we inject proton GCRs based on the following procedure, with a focus on those propagating into (downward) the flux tube while disregarding those moving away (upward) from it. We define $\theta_0$ and $\phi_0$ as the initial polar and azimuthal angles, respectively, relative to the vertical direction. We consider a range for $\theta_0$ spanning from $90^\circ$ to $180^\circ$ (all downward directions) with increments of $\Delta\theta_0 = 1^\circ$, and a range for $\phi_0$ spanning from $0^\circ$ to $360^\circ$ with increments of $\Delta\phi_0 = 45^\circ$. For every combination of values $\left(r_0, \theta_0, \phi_0\right)$, we evaluate $E_p^{\rm k}$ ranging from $1$~GeV to $10^5$~GeV, dividing this range into eight equally spaced logarithmic $E_p^{\rm k}$ bins per decade.

We numerically simulate 3D proton trajectories in the flux tube. For each combination of values $\left(r_0, \theta_0, \phi_0, E_p^{\rm k}\right)$, one proton GCR is injected. The simulation of each proton trajectory is terminated when the proton satisfies one of the following conditions: (1) $z<-1200$~km, (2) $z>1600$~km, or (3) exiting the surface of the tube. In (1), protons are not magnetically reflected before being absorbed by the Sun. We select a minimum height of the simulation box at $-1200$~km where the $pp$ interaction probability exceeds $95\%$. This is to ensure a sufficiently high probability of $pp$ interactions at the bottom of the simulation box. In (2), protons have been reflected and exit the tube from the top cross-section surface, above which the gas density is too low to produce significant gamma rays. In (3), protons pass through the edge of the tube; those with downward velocities later enter the internetwork regions.

\subsubsection{Injection into the Flux Sheet} \label{subsubsec:injection_sheet}

We apply a similar injection process at the top cross-section of the flux sheet at $z=600$~km (see Figure~\ref{fig:schematic_diagram}(b) and Figure~\ref{fig:model_schematics}). We only inject proton GCRs along the $y$ direction, as $\vect{B}$ is invariant in the $x$ direction. Starting at $y_0=20$~km, we increment $y_0$ by $40$~km until $y_0$ reaches the boundary of the flux sheet.

We maintain the same parameter values for $\theta_0$, $\phi_0$, and $E_p^{\rm k}$ as in the flux tube. We use $\tilde{\theta}_0$ and $\tilde{\phi}_0$ to denote the initial polar and azimuthal angles at the injection site of the flux sheet. For each combination of values $\left(y_0, \tilde{\theta}_0, \tilde{\phi}_0, E_p^{\rm k}\right)$, one proton is injected. The simulation of each proton trajectory is terminated when the proton satisfies one of the following conditions: (1) $z<-1200$~km, (2) $z>600$~km, or (3) exiting the surface of the sheet.

We note that the simulation procedure for the flux sheet is independent of the injected GCR spectrum. However, the injected GCR spectrum for the flux sheet is anticipated to be different from the injected GCR spectrum for the flux tube. This is because low-energy proton GCRs are confined within the flux tube, whereas high-energy GCRs traverse the tube with a slight angular deflection, as described in the double-structure model explanation found in Section~\ref{subsec:model_schematic}. Therefore, only the high-energy GCRs can exit the flux tube and subsequently enter the flux sheet. To address this effect, we separately compute the distribution for GCR injection into the flux sheet in Section~\ref{subsubsec:exit_fraction}.

\subsubsection{Angular and Energy Efficiency of GCRs Injected into the Flux Sheet} \label{subsubsec:exit_fraction}

In the simulation of proton GCR injection into the flux tube presented in Section~\ref{subsubsec:injection_tube}, we record the ones that pass through the flux tube surface at heights above $z=0~{\rm km}$. For each $r_0$ and $E_p^{\rm k}$, we record the number of proton GCRs, $Q$, and their polar angles $\theta^\prime$ (relative to the $\vect{\hat{z}}$ direction) upon exiting the tube surface. We disregard the azimuthal angular dependence in $Q$, assuming each proton enters the flux sheet isotropically in the azimuthal direction at the injection site. To account for the higher number of protons injected at larger $r_0$, we weight $Q$ over the ring area with the radius $r_0$ and the width $dr_0$, i.e.,
\beq
\baln
    \langle f\left(\theta^\prime, E_p^{\rm k}\right) \rangle \equiv \frac{\Delta\phi_0}{2\pi} \int_0^{R_{\rm top}} Q\left(r_0, \theta^\prime, E_p^{\rm k}\right) \frac{2\pi r_0}{A_{\rm tot}} dr_0,
\ealn
\eeq
where $A_{\rm tot} = \pi R_{\rm top}{}^2$. Note that if there were no magnetic field within the flux tube, we would have $\langle f\left(\theta^\prime, E_p^{\rm k}\right) \rangle \to 1$ for all values of $\theta^\prime$ and $E_p^{\rm k}$.

Here, we clarify the reason for recording the proton GCRs that pass through the flux tube surface at heights above $z=0$~km instead of above $z=600$~km to obtain $\langle f\left(\theta^\prime, E_p^{\rm k}\right) \rangle$. First, due to the comparatively weaker magnetic field strength in the network field within the lower coronal and chromosphere regions, it is anticipated that the majority of high-energy proton GCRs will already traverse the network fields at heights far above $z=1600$~km and enter the internetwork regions (Figure~\ref{fig:schematic_diagram}(b)). In other words, we anticipate that $\langle f\left(\theta^\prime, E_p^{\rm k}\right) \rangle \to 1$ for $E_p^{\rm k}$ above a certain threshold that needs to be determined. However, in our model, the injection of proton GCRs into the flux tube occurs at $z=1600$~km, which is already close to the photospheric surface. At this height, the flux tube area is $A_{\rm top} = 5.89\times 10^5~{\rm km^2}$. Now, if we choose $z=600$~km as the criteria for recording the proton GCRs passing the flux tube surface, the flux tube area at this height is $1.96\times 10^5~{\rm km^2} \approx 33\% \times A_{\rm top}$. This choice would not result in $\langle f\left(\theta^\prime, E_p^{\rm k}\right) \rangle \to 1$ because proton GCRs with initial injection polar angle $\theta_0 \gtrsim 155^\circ$ have not yet passed through the flux tube surface for such a short vertical distance, as from $z=1600$~km to $z=600$~km. (Note that in the ideal scenario where the injection started at heights within the chromosphere or lower coronal region, these proton GCRs with $\theta_0 \gtrsim 155^\circ$ would have already passed through the network field and entered the internetwork regions.) To circumvent this issue, we choose $z=0$ as the criterion for recording the proton GCRs passing the tube surface. At $z=0$~km, the flux tube radius is $2.01\times 10^4~{\rm km^2} \approx 3.4\% \times A_{\rm top}$. For this choice, the proton GCRs injected at $\theta_0$ close to $180^\circ$ would have enough vertical distance, from $z=1600$~km to $z=0$~km, to escape the flux tube, resulting in $\langle f\left(\theta^\prime, E_p^{\rm k}\right) \rangle \to 1$ for large $E_p^{\rm k}$.

\begin{figure}[t] 
   \centering
   \includegraphics[width=0.99\columnwidth]{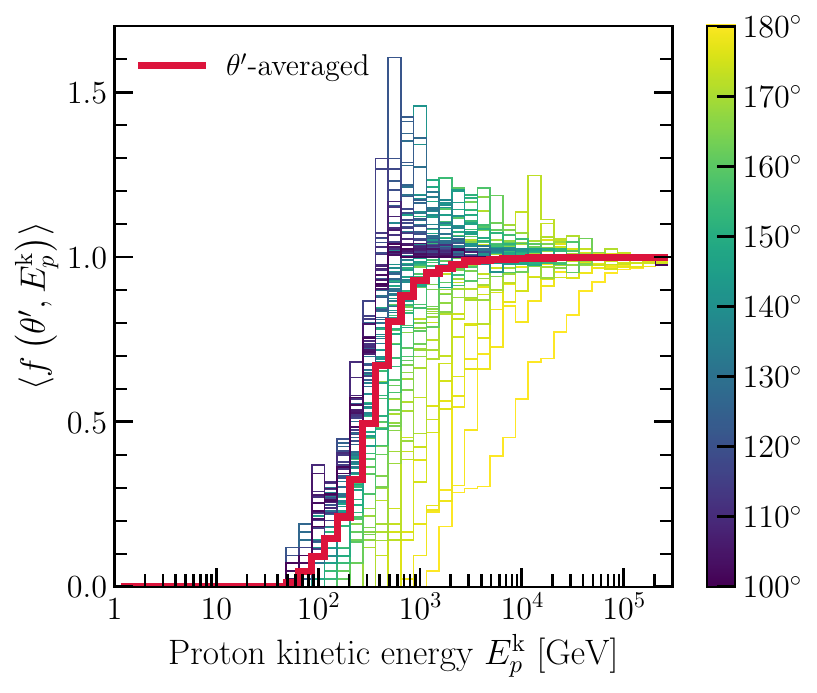}
   \caption{Angular and energy efficiency, $\langle f\left(\theta^\prime, E_p^{\rm k}\right) \rangle$, of proton GCRs passing through the flux tube surface. The color bar corresponds to the polar angle, $\theta^\prime$, at which the proton GCR exits the flux tube surface.}
   \label{fig:frac_exit}
\end{figure}

Figure~\ref{fig:frac_exit} shows the numerical results of $\langle f\left(\theta^\prime, E_p^{\rm k}\right) \rangle$. Each thin colored line represents the polar angle $\theta^\prime$ of a proton GCR upon exiting the flux tube surface. At $E_p^{\rm k}\lesssim 10^2$~GeV, we find $\langle f\left(\theta^\prime, E_p^{\rm k}\right) \rangle \to 0$ because protons injected into the tube are magnetically reflected and do not pass through the tube surface. At $E_p^{\rm k} \gtrsim 10^5$~GeV, we find $\langle f\left(\theta^\prime, E_p^{\rm k}\right) \rangle \to 1$ for all $\theta^\prime$ because protons have such large momenta that the trajectories are not affected by the magnetic fields in the flux tube. At $10^2~{\rm GeV} \lesssim E_p^{\rm k} \lesssim 10^5~{\rm GeV}$, protons are deflected by the magnetic fields and exit the tube surface with smaller polar angles. Because of the deflection, a fraction of proton GCRs entering the tube with different $r_0$, $\theta_0$, and $\phi_0$ can exit the tube surface with nearly the same $\theta^\prime$, leading to certain lines in Figure~\ref{fig:frac_exit} showing $\langle f\left(\theta^\prime, E_p^{\rm k}\right) \rangle > 1$.

In Figure~\ref{fig:frac_exit}, the red thick line represents the average of $\langle f\left(\theta^\prime, E_p^{\rm k}\right) \rangle$ over $\theta^\prime$ ranging from $90^\circ$ to $180^\circ$. It is expressed as
\beq
    \langle f\left(E_p^{\rm k}\right) \rangle \equiv {\int_{90^\circ}^{180^\circ} \langle f\left(\theta^\prime, E_p^{\rm k}\right) \rangle \, d\theta^\prime} \bigg/ {\int_{90^\circ}^{180^\circ} d\theta^\prime},
\eeq
which follows $\langle f\left(E_p^{\rm k}\right) \rangle \leq 1$. We can understand $\langle f\left(E_p^{\rm k}\right) \rangle$ as the fraction of the total numbers of injected protons with $E_p^{\rm k}$ capable of passing through the flux tube surface. It reveals that within the range of $10^2~{\rm GeV} \lesssim E_p^{\rm k} \lesssim 10^3~{\rm GeV}$, which corresponds to the \emph{rising part} of the red line, a fraction of the injected protons traverse through the flux tube, albeit with a slight angular deflection. For $E_p^{\rm k} \gtrsim 10^3$~GeV, protons injected into the flux tube are not captured by the flux tube, but all pass through the tube surface.

Last, we assume that proton GCRs exiting the tube surface follow straight-line trajectories (i.e., no magnetic fields) until they reach the top of the flux sheet. As a result, we have $\langle f\left(\theta^\prime, E_p^{\rm k}\right) \rangle \to \langle f\left(\tilde{\theta}_0, E_p^{\rm k}\right) \rangle$.

\subsection{Gamma-ray Energy Spectra from pp Collisions} \label{subsec:pp_collision}

In this subsection, we follow the methodology of \cite{2006PhRvD..74c4018K} for calculating the gamma-ray energy spectra from $pp$ collisions. They provide an analytical form for the high-energy regime ($E_\gamma \geq 100~{\rm GeV}$) and a $\delta$-function approximation approach for the low-energy regime ($1~{\rm GeV} \lesssim E_\gamma \leq 100~{\rm GeV}$), where they note that this approximation is more accurate in this energy range.

\subsubsection{Higher-energy Regime} \label{subsubsec: high_energy_regime}
For $E_\gamma \geq 100~{\rm GeV}$, we utilize the analytical expression of $F_\gamma$, the gamma-ray energy spectrum resulting from $pp$ interactions, as provided by \cite{2006PhRvD..74c4018K} in their Equation~(58). First, we consider the interaction of a proton GCR with a proton in the solar gas, where the proton GCR has total energy $E_p = E_p^{\rm k} + m_p c^2$. The number of gamma-ray photons per $pp$ interaction within the energy interval $[E_\gamma, E_\gamma + dE_\gamma ]$ is given by
\beq
     dn_\gamma \equiv F_\gamma \left( \frac{E_\gamma}{E_p}, E_p \right) \frac{dE_\gamma}{E_p} .
     \label{eq:gamma_spectrum_pp}
\eeq

The analytical expression of $F_\gamma$ provided in \cite{2006PhRvD..74c4018K} is a parameterization of the gamma-ray energy spectrum obtained from the numerical simulation performed with the \texttt{SIBYLL} code \citep{1994PhRvD..50.5710F}, which takes into account inelastic $pp$ interactions and the subsequent decays of the secondary $\pi^0$ and other mesons into gamma rays. They have shown that their result maintains an accuracy within a few percent under conditions where $E_\gamma/E_p \geq 10^{-3}$ and $E_p > 100$~GeV. Because of the second condition, we will use $F_\gamma$ in their Equation~(58) to calculate the solar disk gamma-ray emission spectrum, $dN_\gamma /dE_\gamma$, from the solar disk for $E_\gamma \geq 100$~GeV.

\subsubsection{Lower-energy Regime} \label{subsubsec: low_energy_regime}
For $1~{\rm GeV} \leq E_\gamma \leq 100~{\rm GeV}$, we utilize the $\delta$-function approximation provided in \cite{2006PhRvD..74c4018K} in their Equations~(77) and (78). We show their methodology of the $\delta$-function approximation in the same form as in Equation~\eqref{eq:gamma_spectrum_pp},
\beq
    dn_\gamma \equiv G_\gamma \left( E_p^{\rm k} \right) \frac{dE_\gamma}{E_p^{\rm k}},
\eeq
where $G_\gamma$, denoting the gamma-ray energy spectrum produced in $pp$ interactions under the assumption of the $\delta$-function approximation, is presented as
\beq
    G_\gamma \left(E_p^{\rm k} \right) = \frac{ 2 n_{\rm f} E_p^{\rm k} }{\sqrt{ \left(K_\pi E_p^{\rm k}\right)^2 - m_\pi^2 }}.
\eeq
Here, the parameter $n_{\rm f}$ is selected as a free parameter to match $dN_\gamma /dE_\gamma$ from the $\delta$-function approximation for $1~{\rm GeV} \leq E_\gamma \leq 100~{\rm GeV}$ and $dN_\gamma /dE_\gamma$ from the analytical expression for $E_\gamma \geq 100~{\rm GeV}$ at $E_\gamma = 100$~GeV. The parameter $K_\pi$ represents the mean fraction of $E_p^{\rm k}$ transferred to the secondary $\pi^0$ and other mesons. The value of $K_\pi = 0.17$ provides a satisfactory agreement with Monte Carlo simulations based on laboratory data, as discussed in \cite{2006PhRvD..74c4018K} and \cite{2000A&A...362..937A}. The $\delta$-function approximation is valid when $\sigma_{pp}$ is nearly constant, which occurs for $E_p \gtrsim 3 E_{\rm th} = 3.66~{\rm GeV}$ where $E_{\rm th} = \left(m_p + m_\pi + m_\pi^2/ 2 m_p\right) c^2 = 1.22~{\rm GeV}$ is the threshold energy of $\pi^0$ production with $m_\pi$ being the rest mass of $\pi^0$ \citep{2006PhRvD..74c4018K}. In this work, while we present our numerical results for $dN_\gamma /dE_\gamma$ down to $E_\gamma = 1$~GeV, we consider the validity of the results to be limited to $E_\gamma \gtrsim 3.66~{\rm GeV}$.

Last, $F_\gamma$ from the analytical expression and $G_\gamma$ from the $\delta$-function approximation method given in \cite{2006PhRvD..74c4018K} do not cover the angular profile of secondary gamma rays from ${pp}$ interactions. As a result, $F_\gamma$ and $G_\gamma$ should be used under the assumption that secondary gamma rays are collinear with the primary proton. This assumption is applied for calculating solar disk gamma-ray emission in Section~\ref{subsec:gamma_ray_emission_highE}. In Appendix~\ref{appendix:collinearity}, we demonstrate that considering a nonzero emission angle would only enhance the total solar disk gamma-ray spectrum by $\approx 3\%$ at $E_\gamma\sim 1$~GeV. Therefore, the assumption of collinearity is adequate for this work.

\subsection{Gamma-ray Emission for Higher-energy Regime} \label{subsec:gamma_ray_emission_highE}

In this subsection, we calculate solar disk gamma-ray spectra from the flux tube and the flux sheet for the higher-energy regime ($E_\gamma \geq 100~{\rm GeV}$).

\emph{Flux tube}:~For proton GCRs injected into the tube (``tb'') at an axial distance $r_0$ and the injection height $z=1600$~km, the resulting gamma-ray flux in the gamma-ray energy interval $[E_\gamma, E_\gamma + dE_\gamma]$ is given by
\beq
\baln
    \frac{dN_{\rm \gamma, tb} \left(r_0\right)}{dE_\gamma} \bigg\lvert_{R_\odot} &= \int_{\Omega_0}  \int_{E_\gamma}^{100 \, E_\gamma} F_\gamma\left(\frac{E_\gamma}{E_p}, E_p\right) \Phi_{p}\left(E_p^{\rm k}\right) \\  
    &\cos\theta_0 \, \mathcal{S}_p\left(r_0, \theta_0, \phi_0, E_p^{\rm k}\right) \, \frac{dE_p}{E_p} \, d\Omega_0,
    \label{eq:gamma_flux_solar_surface_r0}
\ealn
\eeq
where the subscript $R_\odot$ denotes this flux is evaluated at the solar surface. The function $\Phi_p\left(E_p^{\rm k}\right)$ represents the differential proton GCR flux per steradian per proton kinetic energy interval [$E_p^{\rm k}$, $E_p^{\rm k}+dE_p^{\rm k}$] at the injection site of the flux tube. We use the GCR flux at $1~{\rm au}$ measured by the AMS-02 and CREAM (see Section~\ref{subsec:solar_modulation}). The factor $\cos\theta_0$ accounts for the effective flux of proton GCRs entering the horizontal tube cross-section surface at the injection height ($z=1600$~km), and $d\Omega_0 = \sin\theta_0 d\theta_0 d\phi_0$ is the differential solid angle for injection into the flux tube. The factor $dE_p/E_p$ is a product of $1/E_p$ from Equation~\eqref{eq:gamma_spectrum_pp} and the differential integration variable $dE_p$ used to integrate $\Phi_p$. For the upper limit on $E_p$, we use $100\, E_\gamma$. The contribution from $E_p > 100 \, E_\gamma$ to the gamma-ray flux is less than $1\%$ and can be neglected.

The function $\mathcal{S}_p$ represents the integrated absorption probability along the particle trajectory for a proton GCR injected into the tube with initial conditions $\left(r_0, \theta_0, \phi_0, E_p^{\rm k}\right)$, and is furthermore weighted by the gamma-ray transmission factor, $\zeta\left(\vect{r}\right)$, for a gamma ray produced at location $\vect{r}$, thereby taking into account the fraction of gamma rays transmitted outward from the Sun. We express $\mathcal{S}_p$ as
\beq
\baln
    \mathcal{S}_{p}\left(r_0, \theta_0, \phi_0, E_p^{\rm k}\right) = \int_0^{\chi_p^{\rm max}} \zeta\left(\vect{r}\right) \frac{dP_{\rm abs}\left(\chi_p, E_p^{\rm k}\right)}{d\chi_p} \, d\chi_p .
    \label{eq:Z_p_tube}
\ealn
\eeq
Here, $\chi_p = \chi_p\left(\vect{r}\right)$ is the integrated column density of a proton GCR along its trajectory from the injection site $\vect{r}_{\rm inj}$ to the location $\vect{r}$ in hydrogen gas,
\beq
    \chi_p\left(\vect{r}\right) = \int_{\vect{r}_{\rm inj}}^{\vect{r}} \frac{X_{\rm H}\left(z^\prime\right)  \rho\left(z^\prime\right)}{m_{\rm H}} \, \lvert d \vect{r}^\prime\rvert ,
\eeq
where $m_{\rm H}$ is the mass of a hydrogen atom. The upper bound of the integral, $\chi_p^{\rm max}$, in Equation~\eqref{eq:Z_p_tube} denotes the total integrated column density throughout the particle trajectory, from the injection site $\vect{r}_{\rm inj}$ to the exit point $\vect{r}_{\rm exit}$, i.e., $\chi_p^{\rm max} = \chi_p\left(\vect{r}_{\rm exit}\right)$. The function $P_{\rm abs}\left(\chi_p, E_p^{\rm k}\right)$ in Equation~\eqref{eq:Z_p_tube} is the absorption probability of a proton GCR with the accumulated column density $\chi_p$,
\beq
    P_{\rm abs}\left(\chi_p, E_p^{\rm k}\right) = 1 - \exp \left[ {-\chi_p \sigma_{\rm inel}\left(E_p\right)} \right],
\eeq
where $\sigma_{\rm inel}\left(E_p\right)= \sigma_{\rm inel}\left(E_p^{\rm k} + m_p c^2\right)$ is the cross-section of the inelastic $pp$ interaction between the proton GCR and a hydrogen atom in the solar atmosphere. We use the numerical fit of $\sigma_{\rm inel}\left(E_p\right)$ presented in \cite{2006PhRvD..74c4018K} in their Equation~(73),
\beq
\baln
    \sigma_{\rm inel}\left(E_p\right) = &\left( 34.3 + 1.88 D
    + 0.25 D^2 \right)  \\
    &\times \Bigg[1-\left(\frac{E_{\rm th}}{E_p}\right)^4\Bigg]^2~{\rm mb},
\ealn
\eeq
where $D = \ln \left( E_p/{\rm 10^3~GeV} \right)$ and $1~{\rm mb} = 10^{-27}~{\rm cm^2}$.

The gamma-ray transmission factor $\zeta\left(\vect{r}\right)$ introduced in Equation~\eqref{eq:Z_p_tube} denotes the probability for a gamma ray photon produced at $\vect{r}$ from the $pp$ interaction being transmitted through the solar gas. It is expressed as
\beq
    \zeta\left(\vect{r}\right) = \exp \left( - \frac{t_{\gamma} \left(\vect{r}\right)}{\lambda_\gamma} \right),
    \label{eq:gamma_transmission_prob}
\eeq
where $\lambda_{\gamma}$ is the photon mass attenuation length. For $E_\gamma \gtrsim 1$~GeV, $\lambda_{\gamma}$ is approximately $80~{\rm g/cm^2}$ for both hydrogen and helium gases \citep{2022PTEP.2022h3C01W}. Here, $t_\gamma \left(\vect{r}\right)$ is the mass column density of the gamma-ray photon produced at $\vect{r}$, which propagates to a far distance, $\vect{r}_{\rm infty}$, from the Sun, i.e.,
\beq
    {t}_{\gamma} \left(\vect{r}\right) = \int_{\vect{r}}^{\vect{r}_{\rm infty}} \rho\left(\vect{r}^\prime\right) \lvert d \vect{r}^\prime\rvert.
    \label{eq:gamma_integrated_colmn}
\eeq
In our numerical calculations of $t_\gamma$, we have incorporated the curvature of the Sun and the $z$-dependence of $\rho$. We assume that the gamma-ray momentum is collinear with the primary proton GCR momentum during each $pp$ interaction. Consequently, the gamma-ray trajectory used in $t_{\gamma} \left(\vect{r}\right)$ in Equation~\eqref{eq:gamma_integrated_colmn} is a straight line beginning at location $\vect{r}$ and is collinear with the velocity vector of the primary proton GCR at location $\vect{r}$. In Appendix~\ref{appendix:collinearity}, we discuss the validity of the collinearity assumption.

The gamma-ray flux in Equation~\eqref{eq:gamma_flux_solar_surface_r0} only represents the case where proton GCRs are injected at $r_0$. To factor in the cross-section surface area at the injection height, Equation~\eqref{eq:gamma_flux_solar_surface_r0} is weighted over the ring area $2\pi r_0 dr_0/A_{\rm tot}$. It is expressed as 
\beq
    \frac{d\overline{N}_{\rm \gamma, tb}}{dE_\gamma} \bigg\lvert_{R_\odot} = \int_{0}^{R_{\rm top}} \frac{dN_{\rm \gamma, tb} \left(r_0\right)}{dE_\gamma} \bigg\lvert_{R_\odot} \, \frac{2\pi r_0}{A_{\rm tot}} \, dr_0
    \label{eq:gamma_flux_tube_solar_surface_avg}
\eeq
where $A_{\rm tot} = \pi R_{\rm top}{}^2$.

\emph{Flux sheet}: For proton GCRs injected into the sheet (``sh'') at distance $y_0$ at the injection height $z=600$~km, the resulting gamma-ray flux in the energy interval $[E_\gamma, E_\gamma + dE_\gamma]$ is given by
\beq
\baln
    \frac{dN_{\rm \gamma, sh} \left(y_0\right)}{dE_\gamma} \bigg\lvert_{R_\odot} = &\frac{W_{\rm sh}}{L} \, \int_{\tilde{\Omega}_0}  \int_{E_\gamma}^{100 \, E_\gamma}  F_\gamma\left(\frac{E_\gamma}{E_p}, E_p\right) \\ 
    &\Phi_{p}\left(E_p^{\rm k}\right)   
    \langle f \left(\tilde{\theta}_0, E_p^{\rm k}\right) \rangle \cos \tilde{\theta}_0 \\ &\mathcal{S}_p\left(y_0, \tilde{\theta}_0, \tilde{\phi}_0, E_p^{\rm k}\right) \, \frac{dE_p}{E_p} \, d\tilde{\Omega}_0,
    \label{eq:gamma_flux_sheet_solar_surface_y0}
\ealn
\eeq
where much is as above in Equation~\eqref{eq:gamma_flux_solar_surface_r0}. The component $\langle f \left(\tilde{\theta}_0, E_p^{\rm k}\right) \rangle$ is the angular and energy efficiency of protons injected into the flux sheet, as shown in Figure~\ref{fig:frac_exit}. It accounts for the effect that those high-energy protons not confined by the flux tube magnetic fields can traverse through the flux tube and subsequently enter the flux sheet. The component $W_{\rm sh}/L$ takes into account the ratio of the flux sheet cross-section area at $z=600$~km to the area between the two adjacent granule lanes, as discussed in the model schematic diagram in Section~\ref{subsec:model_schematic}. We take the separation of the granule lanes to be the mean size of a granule, $L = 1200$~km.

To factor in the cross-section surface of the flux sheet at the injection height, Equation~\eqref{eq:gamma_flux_sheet_solar_surface_y0} is weighted over $dy_0/y_{\rm top}$. It is expressed as
\beq
    \frac{d\overline{N}_{\rm \gamma, sh}}{dE_\gamma} \bigg\lvert_{R_\odot} = \int_{0}^{y_{\rm top}} \frac{dN_{\rm \gamma, sh} \left(y_0\right)}{dE_\gamma} \bigg\lvert_{R_\odot} ~ \frac{dy_0}{y_{\rm top}}.
    \label{eq:gamma_flux_sheet_solar_surface_avg}
\eeq

Finally, the averaged gamma-ray flux evaluated at the solar surface is the sum of gamma-ray fluxes from the flux tube in Equation~\eqref{eq:gamma_flux_tube_solar_surface_avg} and from the flux sheet in Equation~\eqref{eq:gamma_flux_sheet_solar_surface_avg}. By ``averaged,'' we mean that all patches on the surface are considered equivalent due to our assumption of the spherical symmetry of the solar surface. As a result, the averaged gamma-ray flux evaluated at 1~au from the Sun is given as
\beq
    \frac{dN_\gamma}{dE_\gamma} = \left(\frac{R_\odot}{\rm 1~au}\right)^2~\left( \frac{d\overline{N}_{\rm \gamma, tb}}{dE_\gamma} \bigg\lvert_{R_\odot} + \frac{d\overline{N}_{\rm \gamma, sh}}{dE_\gamma} \bigg\lvert_{R_\odot} \right).
    \label{eq:gamma_flux_at_earth}
\eeq

\subsection{Gamma-ray Emission for the Lower-energy Regime} \label{subsec:gamma_ray_emission_lowE}

In this subsection, we present the formulae for solar disk gamma-ray spectra from the flux tube and the flux sheet in the low-energy regime ($1~{\rm GeV} \lesssim E_\gamma \leq 100~{\rm GeV}$).

\emph{Flux tube}: For proton GCRs injected into the flux tube, the gamma-ray flux is given by
\beq
\baln
    \frac{dN_{\rm \gamma, tb} \left(r_0\right)}{dE_\gamma} \bigg\lvert_{R_\odot} &= \int_{\Omega_0}  \int_{E_{\rm min}/{K_\pi}}^{100 \, E_\gamma} G_\gamma\left(E_p^{\rm k}\right) \Phi_{p}\left(E_p^{\rm k}\right) \\  
    &\cos\theta_0 \mathcal{S}_p\left(r_0, \theta_0, \phi_0, E_p^{\rm k}\right) \, \frac{dE_p^{\rm k}}{E_p^{\rm k}} \, d\Omega_0,
    \label{eq:gamma_flux_solar_surface_r0_lowE}
\ealn
\eeq
where $E_{\rm min} = E_\gamma + m_\pi^2 c^4/4E_\gamma$ \citep{2006PhRvD..74c4018K}. The free parameter $n_{\rm f}$ in $G_r$ in Equation~\eqref{eq:gamma_flux_solar_surface_r0_lowE} for the low-energy regime needs to be adjusted to $2.71$ to match ${dN_{\rm \gamma, tb} \left(r_0\right)} / {dE_\gamma} \lvert_{R_\odot}$ in Equation~\eqref{eq:gamma_flux_solar_surface_r0} for the high-energy regime.

\emph{Flux sheet}: For proton GCRs injected into the flux sheet, the gamma-ray flux is given by
\beq
\baln
    \frac{dN_{\rm \gamma, sh} \left(y_0\right)}{dE_\gamma} \bigg\lvert_{R_\odot} = \, &\frac{W_{\rm sh}}{L} \, \int_{\tilde{\Omega}_0}  \int_{{E_{\rm min}}/{K_\pi}}^{100 \, E_\gamma}  G_\gamma\left(E_p^{\rm k}\right) \\ 
    &\Phi_{p}\left(E_p^{\rm k}\right)   
    \langle f \left(\tilde{\theta}_0, E_p^{\rm k}\right) \rangle \cos \tilde{\theta}_0 \\ &\mathcal{S}_p\left(y_0, \tilde{\theta}_0, \tilde{\phi}_0, E_p^{\rm k}\right) \, \frac{dE_p^{\rm k}}{E_p^{\rm k}} \, d\tilde{\Omega}_0.
    \label{eq:gamma_flux_sheet_solar_surface_y0_lowE}
\ealn
\eeq
The free parameter $n_{\rm f}$ in $G_r$ in Equation~\eqref{eq:gamma_flux_sheet_solar_surface_y0_lowE} for the low-energy regime needs to be adjusted to $1.23$ to match ${dN_{\rm \gamma, sh} \left(y_0\right)} / {dE_\gamma} \lvert_{R_\odot}$ in Equation~\eqref{eq:gamma_flux_sheet_solar_surface_y0} for the high-energy regime.

The methodology for calculating the averaged gamma-ray flux at 1~au from the Sun for the low-energy regime follows the same approach as outlined in Section~\ref{subsec:gamma_ray_emission_highE} for the high-energy regime. As a result, we do not repeat the formulae here.

\begin{figure*}[t] 
   \centering
   \includegraphics[width=0.86 \textwidth]{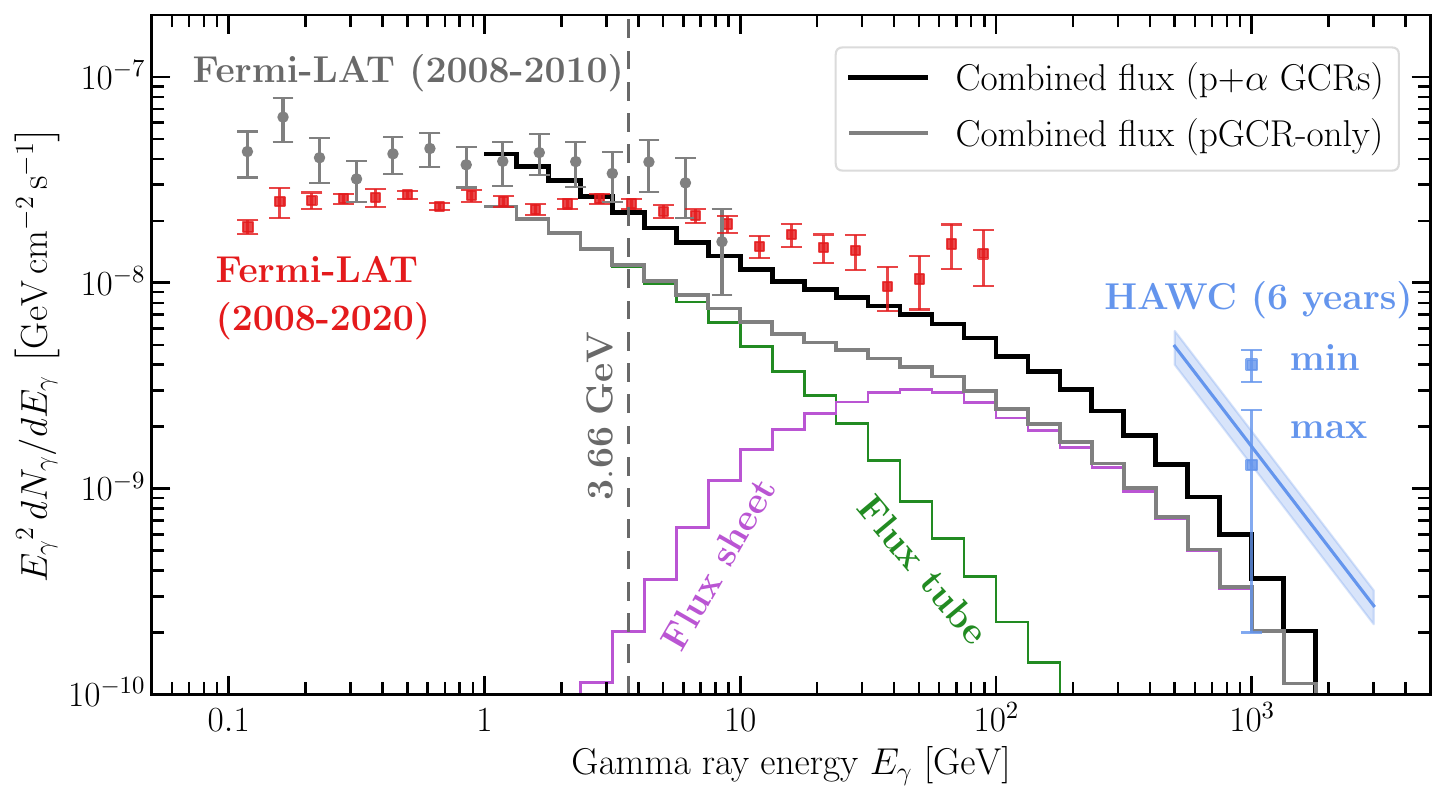}
   \caption{Gamma-ray spectrum of the solar disk. Green and magenta lines are our calculated gamma-ray spectra for the flux tube and sheet due to proton GCR interactions, with the gray line denoting the combined flux. Black line is an $80\%$ increase over the gray line, factoring in helium GCR interactions. Vertical dashed line at $3.66~{\rm GeV}$ is the lowest $E_\gamma$ for this work to be valid. Also presented are Fermi-LAT data from the solar minimum (2008-2010) \citep{2011ApJ...734..116A} and over the full cycle (2008-2020) \citep{2022PhRvD.105f3013L}, in addition to HAWC data from the solar maximum (``max'', 2014-2017) and minimum (``min'', 2018-2021) \citep{2023PhRvL.131e1201A}. Blue line and the shaded band are the best-fit and statistical uncertainties of HAWC's 6.1-year (2014-2021) data. The band represents a single energy bin of 0.5--2.6~TeV; the blue data points indicate how its height varies over the solar cycle.}
   \label{fig:gamma_ray_flux_1au_plus_measurements}
\end{figure*}

\subsection{Nuclear Enhancement Factor} \label{subsec:enhancement_factor}

Gamma-ray production is not solely dependent on protons; helium nuclei, or alpha particles, also play a relevant role. In fact, they make up $\simeq 10\%$ of the number abundance in both GCRs and photospheric gas. To factor in the effect of helium nuclei, we utilize the result of nuclear enhancement factor $\varepsilon_{\rm M}$ from \cite{2014ApJ...789..136K}, which gives results for a variety of cases. Following \cite{2017PhRvD..96b3015Z}, which makes choices based on recent cosmic-ray data, we take $\varepsilon_{\rm M} = 1.8$. This means that the total gamma-ray spectrum should be that much larger than the one calculated using only the proton densities in the Sun and the GCRs.

\section{PREDICTED GAMMA-RAY SPECTRUM} \label{sec:results}

Figure~\ref{fig:gamma_ray_flux_1au_plus_measurements} shows our numerical results of gamma-ray energy spectra, $E_\gamma{}^2 dN_\gamma/dE_\gamma$, over an energy range of $1~{\rm GeV}$ to $2\times 10^3~{\rm GeV}$. The green line is our model prediction of the gamma-ray energy spectrum from the flux tube, and the magenta is from the flux sheet. Both spectra result from proton GCR interactions with hydrogen gas. For $E_\gamma \lesssim 10~{\rm GeV}$, the gamma rays from the flux tube dominate, while the gamma rays from the flux sheet dominate at $E_\gamma \gtrsim {\rm few}\times 10~{\rm GeV}$. This shift arises because lower-energy proton GCRs ($E_p^{\rm k} \lesssim 10^2~{\rm GeV}$) are magnetically confined within the flux tube. In contrast, the higher-energy proton GCRs ($E_p^{\rm k} \gtrsim {\rm few}\times 10^2~{\rm GeV}$) are capable of traversing through the flux tube and entering the flux sheet, as indicated from the angular and energy efficiency in Figure~\ref{fig:frac_exit}. Furthermore, due to the higher magnetic field strength at the injection height of the flux sheet compared to the flux tube, particle reflection is more effective, consequently leading to increased gamma-ray yields in the flux sheet.

In Figure~\ref{fig:gamma_ray_flux_1au_plus_measurements}, the gray line is the combined gamma-ray energy spectrum summing the proton interactions in the flux tube and flux sheet. To account for the helium GCR interactions discussed in Section~\ref{subsec:enhancement_factor}, we multiply the gray line by the nuclear enhancement factor, $\varepsilon_{\rm M} = 1.8$, resulting in the black line. Consequently, the black line represents the total gamma-ray energy spectrum from the solar disk. The black line shows a modest spectral slope, $dN_\gamma/dE_\gamma \sim E_\gamma^{-2.4}$, for $1~{\rm GeV} \lesssim E_\gamma \lesssim 10^2~{\rm GeV}$, aligning close to the spectral shape from Fermi-LAT observations over the entire solar cycle. This moderate spectral slope is a product of two combined emission sources: the gamma-ray emission from the flux tube for $E_\gamma \lesssim 10$~GeV and the emission from the flux sheet for $10~{\rm GeV} \lesssim E_\gamma \lesssim 10^2~{\rm GeV}$. At much higher $E_\gamma$, the spectrum steepens to $dN_\gamma/dE_\gamma \sim E_\gamma^{-3.6}$, agreeing with the spectral shape from the HAWC observation. This steep slope is due to the limited effectiveness of GCR capture and reflection within the finite-sized flux sheet --- an inherent characteristic of the intermittent field distributions generated by granule convective flows. Additionally, the normalization of the black line is reasonably consistent with the observational data, remaining within a factor of about two, ranging from 1 GeV to $10^3$ GeV. This agreement strongly suggests that the observed gamma-ray spectra are influenced and shaped by the flux tubes in the network elements and the flux sheets in the intergranule lanes.

Based on the discussion above, important conclusions and insights can be drawn from Figure~\ref{fig:gamma_ray_flux_1au_plus_measurements}. 
\begin{enumerate}
\item In a realistic solar environment, the observed gamma-ray spectra in the $\sim \text{1--100}~{\rm GeV}$ range are likely shaped by the combination of multiple finite-sized magnetic flux structures at the photosphere. We note that a simple $P$ to $B$ scaling relation would not reproduce the observed spectra over a wide range of $E_\gamma$.
\item Taking into account the finite-sized magnetic flux structures is key to the considerably softer gamma-ray spectra near $10^3~{\rm GeV}$ from HAWC. This is due to the large fraction of high-energy proton GCRs that plunge through both the flux tube and flux sheet.
\item While the data can be reasonably interpreted through these two basic structures, incorporating additional flux structures and field features could potentially enhance the accuracy of the model. This could include variability in size and average magnetic field strength of flux tubes and flux sheets, non-vertical flux structures, and the magnetic turbulence caused by the convective flow of the granules. Taking these into account could lead to a larger predicted flux in the $10^3~{\rm GeV}$ range.
\end{enumerate}

Building on prior theoretical work on the solar-disk gamma-ray emission \citep{1991ApJ...382..652S, 1992NASCP3137..542S, 2017PhRvD..96b3015Z, 2020MNRAS.491.4852H, 2020PhRvD.101h3011M, 2020arXiv200903888L, 2020APh...11902440G, 2022ApJ...941...86G}, our model goes much further by considering magnetic fields motivated by solar physics data and magnetoconvection simulations. Our approach introduces the concept of finite-sized flux structures into the problem, differentiating between higher- and lower-energy proton GCR behaviors by considering two separate flux structures, calculating 3D particle trajectories, and comparing theoretical predictions to data across three orders of magnitude in gamma-ray energy.

How do our results compare to those of previous work? First, the pioneering work of \cite{1991ApJ...382..652S}, which focused on magnetic mirroring in the magnetic canopy fields, produced predictions that disagreed with measurements by a factor of about 5--10. Moreover, their spectrum was predicted only for gamma-ray energies up to 5~GeV. Several alternative approaches have been proposed in recent years. For instance, \cite{2020PhRvD.101h3011M} and \cite{2020arXiv200903888L} focus on the effects of coronal fields, which extend towards the photosphere. However, their results are about three times smaller than the observational data at around 10~GeV and 10~times smaller at 100~GeV. Most importantly, these models do not align with the new HAWC data in terms of normalization and spectral shape at about $10^3$~GeV. Nevertheless, their findings point to the potential significance of solar coronal fields.

For future work, in Figure~\ref{fig:gamma_ray_flux_1au_plus_measurements}, while we show our theory predictions down to $E_\gamma = 1~{\rm GeV}$, we note that they are considered valid only for $E_\gamma \gtrsim 3.66~{\rm GeV}$. This is because the $\delta$-function approximation provided in \cite{2006PhRvD..74c4018K} is only valid for $E_p \gtrsim 3E_{\rm th} = 3.66~{\rm GeV}$ where $\sigma_{pp}$ is nearly constant. Although the method for $E_\gamma < 3.66~{\rm GeV}$ is currently absent in the literature, this task could potentially be performed using publicly available Monte Carlo codes, for instance, \texttt{FLUKA} \citep{2015_Battistoni_fluka, 2022FrP.....9..705A} and \texttt{GEANT4} \citep{2003NIMPA.506..250A}. Further investigation in this aspect is an interesting direction for future work.

Last, in Appendix~\ref{sec:absorption_result}, we present the most probable injection polar angles for GCRs that produce the escaping gamma rays from the solar disk as $\approx 171^\circ \pm 5^\circ$ for the flux tube and $\approx 135^\circ \pm 15^\circ$ for the flux sheet. In Appendix~\ref{appendix:avg_angle_height}, we perform the averaged emission angle and height of the gamma-ray flux. We find that the emission primarily occurs within a height ranging from $-100~{\rm km}$ to $500~{\rm km}$, encompassing the photosphere and extending $\sim 100$~km into the uppermost convection zone. Moreover, we find an emission angle of $\approx 65^\circ \pm 15^\circ$ relative to the normal direction of the solar surface, indicating that the gamma rays predicted by our model predominantly originate from the outer regions of the solar disk. This range of predicted angles is a result of our exclusive focus on vertical flux structures and the omission of magnetic turbulence in our model. In a more realistic solar surface environment, the presence of non-vertical flux structures with magnetic turbulence would have an impact on particle trajectories, causing deviations from the magnetic bottle effect. This can potentially lead to smaller emission angles and a more uniform distribution of gamma-ray emission across the solar disk. We will explore this in future work.

\section{DISCUSSION AND CONCLUSIONS} \label{sec:conclusion}

In this paper, we present a simple model of solar-surface magnetic fields, aiming to understand the broad outline of how hadronic GCRs shape gamma-ray emission. We have investigated GCR propagation and gamma-ray emission within a simplified solar-surface magnetic field model comprising a flux tube and a flux sheet. In the $\sim \text{1--100~GeV}$ range, where emission is due to both the tube and the sheet, our model produces a hard spectrum ($dN_\gamma/dE_\gamma \sim E_\gamma^{-2.4}$). In the $\sim \text{100--1000~GeV}$ range, where the tail of the sheet emission dominates, our model produces a considerably softer spectrum ($dN_\gamma/dE_\gamma \sim E_\gamma^{-3.6}$) due to the limited effectiveness of GCR capture and reflection by the finite-sized flux sheet. Despite the fact that the model presented in this work does not capture all the complexities of the solar-surface magnetic fields, the reasonable agreement between our model predictions and data (within a factor of 2 over about three orders of magnitude in gamma-ray energy) is encouraging and highlights the critical roles of finite-sized field structures of flux tubes and flux sheets in shaping solar-disk gamma-ray emission.

We emphasize that the model, as presented in this study, remains a work in progress and does not yet offer a conclusive solution to the long-standing issue of solar-disk gamma-ray emission. Given the multi-scale nature of the solar magnetic field, which spans from large-scale heliospheric magnetic fields to small-scale \Alfven wave turbulence and dissipation, incorporating a realistic solar magnetic field model into the current problem is a challenging task and necessitates a stepwise approach. The reasonable agreement with data suggests that the double-structure model with finite-sized magnetic flux structure captures key characteristics of the hadronic GCR reflection at the solar surface and the resulting solar gamma-ray emission. This simplified approach sets the stage for further refined modeling and advanced computational efforts, details of which are elaborated as follows.

In future work, it will be important to go beyond the assumptions here of a simple double-structure model in the quiet photosphere with an assumption of the spherical symmetry. Aspects to investigate include the influences of active regions and coronal-hole open field distributions in GCR transport and gamma-ray emission. Ultimately, we need to evaluate the impact of active regions on GCR transport toward the solar surface. We also need to quantify the relationship between the latitudinal dependence of coronal-hole distribution and the resulting gamma-ray emission. Both directions hold great potential in unraveling the observed anti-correlation between solar activity and the gamma-ray flux \citep{2016PhRvD..94b3004N, 2018PhRvD..98f3019T, 2018PhRvL.121m1103L, 2022PhRvD.105f3013L, 2023PhRvL.131e1201A}.

Future studies should also explore GCR propagation and solar gamma-ray emission within the framework of magnetoconvection in the photosphere and uppermost convection zone. In this work, we focused on the vertical magnetic flux tube and flux sheet in magnetohydrostatic equilibrium with the surrounding gas. The perpendicular field components leading to particle reflection are caused by the compression exerted by the surrounding gas pressure. Eventually, we need to understand whether and how magnetic turbulence resulting from the convection of the granule cells can cause a similar effect and affect GCR propagation. Investigating the impact of turbulence may further elucidate additional magnetic field structures at the photosphere that contribute to the observed gamma-ray spectra.

Solar gamma-ray observations have presented an intriguing new opportunity to investigate the characteristics of magnetic fields in the photosphere. The origin of these magnetic fields, whether arising from the decay of active regions \citep{1987SoPh..110..115S} or local fast dynamo actions in the uppermost convection zone \citep{1999ApJ...515L..39C, 2003ApJ...588.1183C, 2007A&A...465L..43V}, remains uncertain. By refining theoretical models of solar magnetic fields and the transport of GCRs, we have the potential to gain valuable insights into this unresolved matter. In this regard, future research should focus on exploring these issues through magnetoconvection simulations and utilizing high-resolution measurements of small-scale magnetic structures at the solar surface, such as those provided by the Daniel~K. Inouye Solar Telescope \citep{2020SoPh..295..172R}.

\vspace{0.3cm}

We are grateful for stimulating discussions with Ofer Cohen, Federico Fraschetti, Hugh Hudson, Mikhail Malkov, and Eleonora Puzzoni. This work was supported by NASA grant Nos.\ 80NSSC20K1354 and 80NSSC22K0040. J.F.B. was additionally supported by National Science Foundation grant No.\ PHY-2012955. J.T.L. thanks the Lunar and Planetary Laboratory for their hospitality while parts of this manuscript were completed.

\vspace{0.2cm}
\appendix

\section{MOST PROBABLE GCR POLAR INJECTION ANGLES} \label{sec:absorption_result}

In this appendix, we show the most probable injection polar angles for GCRs that produce the gamma rays that escape from the solar disk.

For proton GCR injection into the flux tube, we average  $\mathcal{S}_p$ in Equation~\eqref{eq:Z_p_tube} over the injection area, $2\pi r_0 dr_0$, and the azimuthal injection angle, $\phi_0$,
\beq
\baln
    \langle \mathcal{S}_p \rangle_{\rm tb} &= \langle \mathcal{S}_p \left(\theta_0, E_p^{\rm k} \right) \rangle_{\rm tb} \\
    &\equiv \int_0^{2\pi}\int_0^{R_{\rm top}} \mathcal{S}_p \left(r_0, \theta_0, \phi_0, E_p^{\rm k}\right) \frac{2\pi r_0 dr_0}{A_{\rm tot}} \frac{d\phi_0}{2\pi}.
    \label{eq:Z_p_avg_tb}
\ealn
\eeq 
The term $\langle \mathcal{S}_p \rangle_{\rm tb}$ is interpreted as the averaged absorption probability for a primary proton GCR along the section of its trajectory where the produced gamma rays from $pp$ interactions successfully escape without being absorbed by the Sun.

For proton GCR injection into the flux sheet, $\mathcal{S}_p$ is averaged over the injection width $dy_0$ and the azimuthal injection angle $d\phi_0$. It is expressed as
\beq
\baln
    \langle \mathcal{S}_p \rangle_{\rm sh} &= \langle \mathcal{S}_p \left(\tilde{\theta}_0, E_p^{\rm k} \right) \rangle_{\rm sh} \\
    &\equiv \int_0^{2\pi}\int_0^{y_{\rm top}} \langle f\left(\tilde{\theta}_0, E_p^{\rm k} \right) \rangle \\
    &\quad \mathcal{S}_p \left(y_0, \tilde{\theta}_0, \tilde{\phi}_0, E_p^{\rm k} \right) \frac{dy_0}{y_{\rm top}} \frac{d\tilde{\phi}_0}{2\pi}.
    \label{eq:Z_p_avg_sh}
\ealn
\eeq
where $\langle f\left(\tilde{\theta}_0, E_p^{\rm k}\right) \rangle$ accounts for the angular and energy efficiency of the injected particles into the flux sheet. Note that Equation~\ref{eq:Z_p_avg_sh} accounts for monoenergetic protons with a kinetic energy of $E_p^{\rm k}$, without incorporating any weighting based on the proton GCR spectrum.

\begin{figure}[t] 
   \centering
   \includegraphics[width=0.98\columnwidth]{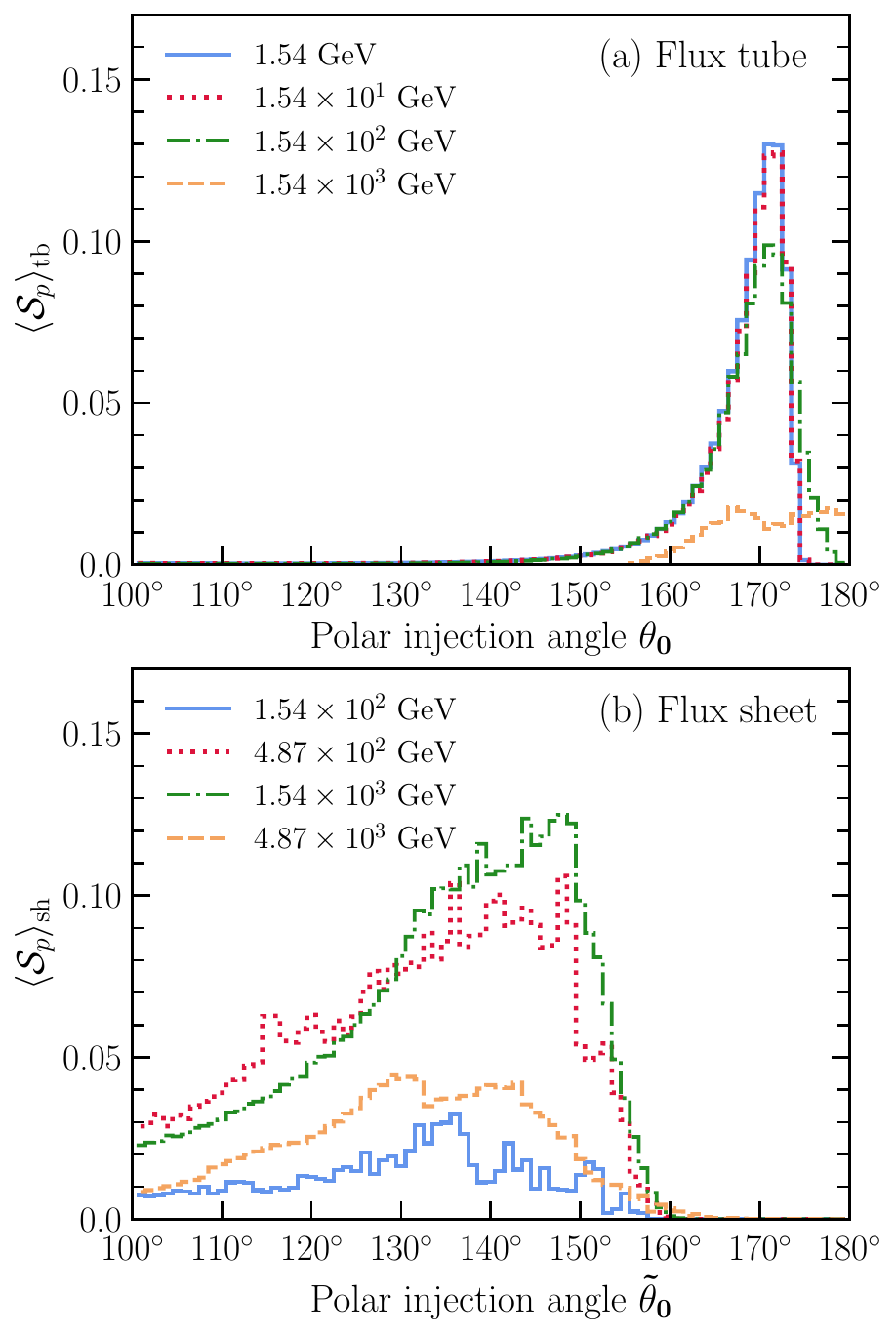}
   \caption{Most probable GCR polar injection angles. The higher the values of $\langle \mathcal{S}_p \rangle$, the greater the production of outward-directed gamma rays. (a) $\langle \mathcal{S}_p \rangle_{\rm tb}$ for the flux tube shown in Equation~\eqref{eq:Z_p_avg_tb}. (b) $\langle \mathcal{S}_p \rangle_{\rm sh}$ for the flux sheet shown in Equation~\eqref{eq:Z_p_avg_sh}.}
   \label{fig:P_abs_reflected_GCR}
\end{figure}

Figure~\ref{fig:P_abs_reflected_GCR}(a) shows our calculations of $\langle \mathcal{S}_p \rangle_{\rm tb}$ in Equation~\eqref{eq:Z_p_avg_tb} for proton GCR injection into the flux tube. The three lowest $E_p^{\rm k}$ lines indicate that the most probable polar injection angle is $\theta_0 \approx 171^\circ \pm 5^\circ$. For $\theta_0 \lesssim 160^\circ$, the injected protons are reflected too high from the solar surface to accumulate significant column density. For $\theta_0 \gtrsim 175^\circ$, the injected protons are not yet reflected by the magnetic fields by the time they reach the bottom of the simulation box; thus, they do not contribute outward-directed gamma rays.

In Figure~\ref{fig:P_abs_reflected_GCR}(a), the three lowest $E_p^{\rm k}$ have approximately equivalent $\langle \mathcal{S}_p \rangle_{\rm tb}$. In contrast, for the highest $E_p^{\rm k}$, then $\langle \mathcal{S}_p \rangle_{\rm tb}$ has substantially smaller values across all most probable $\tilde{\theta}_0$. This difference arises because protons with $E_p^{\rm k} \lesssim 100~{\rm GeV}$ are nearly all reflected within the flux tube via the magnetic bottle effect. In contrast, protons with $E_p^{\rm k} \gtrsim 10^3 \, {\rm GeV}$ do not spiral within the flux tube but pass through the tube surface with only small angular deflection. This result echos the findings from the angular and energy efficiency of GCRs injected into the flux sheet, $\langle f\left(\tilde{\theta}_0, E_p^{\rm k}\right) \rangle$, as presented in Figure~\ref{fig:frac_exit}.

Figure~\ref{fig:P_abs_reflected_GCR}(b) shows our calculations of $\langle \mathcal{S}_p \rangle_{\rm sh}$ in Equation~\eqref{eq:Z_p_avg_sh} for proton GCR injection into the flux sheet. Note that the lines with $E_p^{\rm k} = 1.54~{\rm GeV}$ and $1.54\times 10^1\,{\rm GeV}$ shown earlier in Figure~\ref{fig:P_abs_reflected_GCR}(a) are not shown in Figure~\ref{fig:P_abs_reflected_GCR}(b) because the majority of proton GCRs at such low energy do not penetrate through the flux tube. In Figure~\ref{fig:P_abs_reflected_GCR}(b), all four energies indicate that the most probable injection angle is $120^\circ \lesssim \tilde{\theta}_0 \lesssim 150^\circ$. The two intermediate $E_p^{\rm k}$ show high $\langle \mathcal{S}_p \rangle_{\rm sh}$, indicating that those protons are effectively reflected and produce outward-directed gamma rays. Notably, these two intermediate energies lie within the rising part of $\langle f\left(E_p^{\rm k}\right) \rangle$ in Figure~\ref{fig:frac_exit}. In contrast, the lowest $E_p^{\rm k}$ has lower $\langle \mathcal{S}_p \rangle_{\rm sh}$. This is because the majority of the proton GCRs are magnetically reflected by the flux tube, while only a small fraction of protons successfully penetrate through the flux tube and enter the flux sheet. The highest $E_p^{\rm k}$ also has lower $\langle \mathcal{S}_p \rangle_{\rm sh}$. This is because the efficiency of capturing and reflecting proton GCRs by flux sheet magnetic fields starts diminishing for $E_p^{\rm k} \gtrsim {\rm few \times 10^3~GeV}$. Finally, for $E_p^{\rm k} \gtrsim 10^4\,{\rm GeV}$, $\langle \mathcal{S}_p \rangle_{\rm sh} \approx 0$ across all $\tilde{\theta}_0$, and are not shown in Figure~\ref{fig:P_abs_reflected_GCR}.

In Figure~\ref{fig:P_abs_reflected_GCR}(b), all four energies show $\langle \mathcal{S}_p \rangle_{\rm sh} \approx 0$ for $\tilde{\theta}_0 \gtrsim 160^\circ$. This result implies that protons injected into the flux sheet with $\tilde{\theta}_0 \gtrsim 160^\circ$ do not experience magnetic reflection before being absorbed by the Sun. Therefore, any gamma rays generated from these protons do not contribute to the outward-directed gamma rays. Note that the range of these polar injection angles is significantly broader compared to the flux tube scenario depicted in Figure~\ref{fig:P_abs_reflected_GCR}(a).

\section{AVERAGE EMISSION ANGLE AND HEIGHT} \label{appendix:avg_angle_height}

In this appendix, we present the average emission angle and height of the solar-disk gamma rays. We only present the methodology for the higher-energy regime ($E_\gamma \geq 10^2~{\rm GeV}$). The methodology for the lower-energy regime ($1~{\rm GeV} \lesssim E_\gamma \leq 10^2~{\rm GeV}$), which has been discussed in Section~\ref{subsec:gamma_ray_emission_lowE}, is not repeated here.

\subsection{Average Emission Angle} \label{appendix:avg_angle}

\emph{Flux tube}: First, we calculate the average gamma-ray emission angle in the case of a flux tube. We define $\theta_p\left(\vect{r}\right)$ as the angle of the proton's velocity vector at location $\vect{r}$ relative to the normal direction of the surface. Thus, for proton GCRs injecting into the flux tube at axial distance $r_0$ at the injection height $z=1600$~km, the $\theta_p\left(\vect{r}\right)$-weighted gamma-ray flux is given as
\beq
\baln
    \frac{dN^{\theta_p}_{\rm \gamma, tb} \left(r_0\right)}{dE_\gamma} \bigg\lvert_{R_\odot} &= \int_{\Omega_0}  \int_{E_\gamma}^{100 \, E_\gamma} F_\gamma\left(\frac{E_\gamma}{E_p}, E_p\right) \Phi_{p}\left(E_p\right) \\  
    &\cos\theta_0 \, \mathcal{S}_p^{\theta_p} \left(r_0, \theta_0, \phi_0, E_p^{\rm k}\right) \, \frac{dE_p}{E_p} \, d\Omega_0.
    \label{eq:angle_weighted_flux_r0}
\ealn
\eeq
Here, $\mathcal{S}_p^{\theta_p}$ is defined as $\mathcal{S}_p$ in Equation~\eqref{eq:Z_p_tube} weighted by $\theta_p\left(\vect{r}\right)$, i.e.,
\beq
\baln
    \mathcal{S}_p^{\theta_p} &= \mathcal{S}_p^{\theta_p} \left(r_0, \theta_0, \phi_0, E_p^{\rm k}, E_\gamma\right)\\ &\equiv \int_0^{\chi_p^{\rm max}} \theta_p\left(\vect{r}\right) \zeta\left(\vect{r}\right) \frac{dP_{\rm abs}\left(\chi_p, E_p^{\rm k}\right)}{d\chi_p} \, d\chi_p.
\ealn
\eeq
The $\theta_p$-weighted gamma-ray flux in Equation~\eqref{eq:angle_weighted_flux_r0} is further weighted over the area along the cross-section surface of the flux tube at the injection height,
\beq
    \frac{d\overline{N}^{\theta_p}_{\rm \gamma, tb}}{dE_\gamma} \bigg\lvert_{R_\odot} = \int_{0}^{R_{\rm top}} \frac{dN^{\theta_p}_{\rm \gamma, tb} \left(r_0\right)}{dE_\gamma} \bigg\lvert_{R_\odot} \, \frac{2\pi r_0}{A_{\rm tot}} \, dr_0.
\eeq
As a result, the average emission angle $\langle \theta_p \rangle$ for the gamma-ray flux at $E_\gamma$ is given as 
\beq
    \langle\theta_p\rangle = \langle \theta_p \rangle \left(E_\gamma\right) \equiv {\frac{d\overline{N}^{\theta_p}_{\rm \gamma, tb}}{dE_\gamma} \bigg\lvert_{R_\odot}} \Bigg/ {\frac{d\overline{N}_{\rm \gamma, tb}}{dE_\gamma} \bigg\lvert_{R_\odot}},
    \label{eq:theta_mean}
\eeq
where $d\overline{N}_{\rm \gamma, tb}/ dE_\gamma \lvert_{R_\odot}$ is provided in the calculation of solar-disk gamma-ray flux in Equation~\eqref{eq:gamma_flux_tube_solar_surface_avg}.

Next, we calculate the root mean square (RMS) emission angle, $\theta_p \left(\vect{r}\right) - \langle \theta_p \rangle$, for the gamma-ray flux at $E_\gamma$. This time, the gamma-ray flux is weighted by the mean squared polar angle, $\left( \theta_p \left(\vect{r}\right) - \langle \theta_p \rangle\right)^2$, of each proton's moving direction at location $\vect{r}$. It is given as
\beq
\baln
    \frac{dN^{\Delta\theta_p^2}_{\rm \gamma, tb} \left(r_0\right)}{dE_\gamma} \bigg\lvert_{R_\odot} &= \int_{\Omega_0}  \int_{E_\gamma}^{100 \, E_\gamma} F_\gamma\left(\frac{E_\gamma}{E_p}, E_p\right) \Phi_{p}\left(E_p\right) \\  
    &\cos\theta_0 \, \mathcal{S}_p^{\Delta\theta_p^2} \left(r_0, \theta_0, \phi_0, E_p^{\rm k}\right) \, \frac{dE_p}{E_p} \, d\Omega_0.
    \label{eq:angle_square_weighted_flux_r0}
\ealn
\eeq
where $\mathcal{S}_p^{\Delta\theta_p^2}$ is defined as $\mathcal{S}_p$ in Equation~\eqref{eq:Z_p_tube} weighted by $\left( \theta_p \left(\vect{r}\right) - \langle \theta_p \rangle \right)^2$, i.e.,
\beq
\baln
    \mathcal{S}_p^{\Delta\theta_p^2} &= \mathcal{S}_p^{\Delta\theta_p^2} \left(r_0, \theta_0, \phi_0, E_p^{\rm k}, E_\gamma\right) \\ 
    &\equiv \int_0^{\chi_p^{\rm max}} \left( \theta_p\left(\vect{r}\right) - \langle \theta_p \rangle \right)^2 \zeta \left( \vect{r} \right) \frac{dP_{\rm abs}\left(\chi_p, E_p^{\rm k}\right)}{d\chi_p} \, d\chi_p.
\ealn
\eeq
Equation~\eqref{eq:angle_square_weighted_flux_r0} is further weighted over the area along the cross-section surface of the flux tube at the injection height,
\beq
    \frac{d\overline{N}^{\Delta\theta_p^2}_{\rm \gamma, tb}}{dE_\gamma} \bigg\lvert_{R_\odot} = \int_{0}^{R_{\rm top}} \frac{dN^{\Delta\theta_p^2}_{\rm \gamma, tb} \left(r_0\right)}{dE_\gamma} \bigg\lvert_{R_\odot} \, \frac{2\pi r_0}{A_{\rm tot}} \, dr_0.
\eeq
As a result, the RMS emission angle $\langle \Delta \theta_p \rangle$ for the gamma-ray flux at $E_\gamma$ is given as
\beq
    \langle \Delta \theta_p \rangle = \langle \Delta \theta_p \rangle \left(E_\gamma\right) \equiv \sqrt{ {\frac{d\overline{N}^{\Delta\theta_p^2}_{\rm \gamma, tb}}{dE_\gamma} \bigg\lvert_{R_\odot}} \Bigg/ {\frac{d\overline{N}_{\rm \gamma, tb}}{dE_\gamma} \bigg\lvert_{R_\odot}} }.
\eeq

\emph{Flux sheet}: The methodology for the average and RMS emission angles in the case of the flux sheet is the same as in the flux tube. The only difference is the integrants in Equations~\eqref{eq:angle_weighted_flux_r0} and~\eqref{eq:angle_square_weighted_flux_r0} should be further multiplied by the angular and energy efficiency $\langle f\left(\tilde{\theta}_0, E_p^{\rm k}\right) \rangle$ shown in Figure~\ref{fig:frac_exit}. Consequently, we do not repeat the formulae here.

\begin{figure}[t] 
   \centering
   \includegraphics[width=0.98\columnwidth]{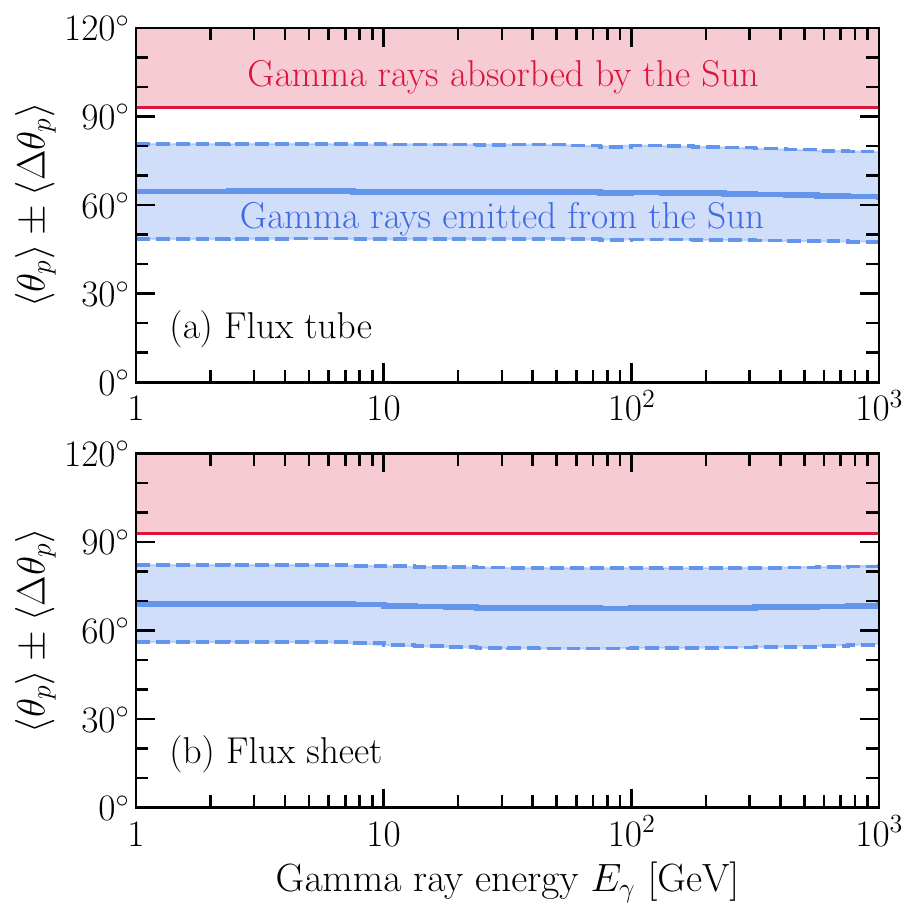}
   \caption{Average (blue lines) and RMS (blue bands) emission angles for gamma rays successfully transmitted from the Sun.}
   \label{fig:mean_emission_angle}
\end{figure}

Figure~\ref{fig:mean_emission_angle} shows the average (blue solid lines) and RMS (blue shaded bands) emission angles in the cases for the flux tube (top) and flux sheet (bottom) for gamma rays successfully transmitted from the Sun. Our result indicates that for each small patch at the surface of the Sun, the average emission angle is $\approx 65^\circ$, relative to the normal direction, with a corresponding RMS emission angle of approximately $15^\circ$. Finally, gamma rays are fully absorbed by the Sun whenever $\theta_p\left(\vect{r}\right) \geq  \theta_{p,\,{\rm crit}} \approx 93^\circ$, which are highlighted in red shaded bands in Figure~\ref{fig:mean_emission_angle}.

\subsection{Average Emission Height} \label{appendix:avg_height}

The methodology employed to calculate the average and RMS emission heights (denoted as $\langle z_p \rangle$ and $\langle \Delta z_p \rangle$, respectively) are consistent with the approach described in Appendix~\ref{appendix:avg_angle} for computing emission angles. Therefore, we do not reiterate the formulas here.

\begin{figure}[t] 
   \centering
   \includegraphics[width=0.98\columnwidth]{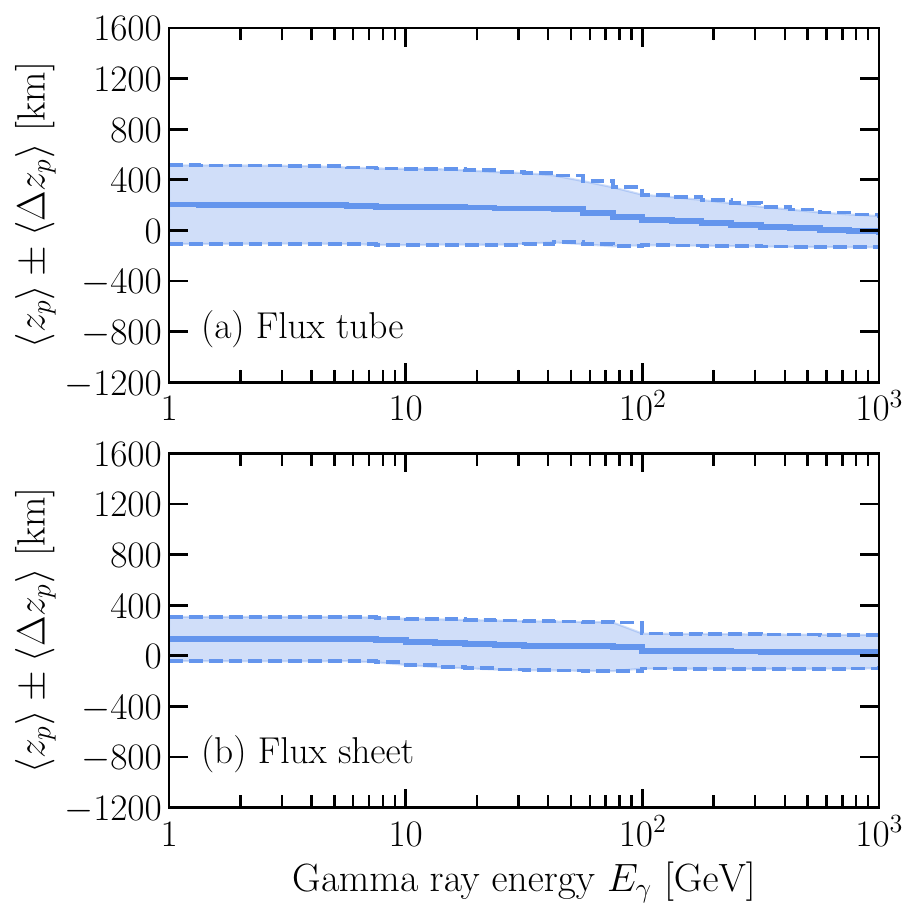}
   \caption{Average (black lines) and RMS (blue bands) emission heights for gamma rays successfully transmitted from the Sun.}
   \label{fig:mean_emission_height}
\end{figure}

Figure~\ref{fig:mean_emission_height} shows $\langle z_p \rangle$ and $\langle \Delta z_p \rangle$ for the gamma rays produced in the flux tube (top) and flux sheet (bottom). Our findings indicate that emission typically occurs at the height of $z \approx 0$~km with an RMS of $\approx 300$~km for the flux tube and with an RMS of $\approx 150$~km for the flux sheet. Consequently, this suggests that the layers responsible for producing the gamma rays successfully transmitted from the Sun occur in the photosphere and a thin layer extending $\sim 100$~km into the upper convection zone.

\section{IMPACT OF NON-COLLINEAR EMISSION ON GAMMA-RAY FLUX} \label{appendix:collinearity}

In this appendix, we estimate the errors incurred by assuming that gamma rays are collinear with their parent protons.

In this simplified model, we consider a gamma-ray emission cone with a half-angle denoted as $\Delta\Theta_\gamma$. Accordingly, the solid angle, $\Omega_\gamma$, of the emission cone is expressed as $2 \pi \left(1 - \cos{\Delta\Theta_\gamma}\right)$. We assume that the number of gamma-ray photons distribute uniformly in the angular direction in $\Omega_\gamma$. Under this assumption, we conduct a numerical calculation of $\mathcal{S}_p$ in Equation~\eqref{eq:Z_p_tube} to evaluate the impact of a nonzero emission cone angle on the solar gamma-ray flux. (To remind the reader, $\mathcal{S}_p$ in Equation~\eqref{eq:Z_p_tube} is the total proton GCR absorption probability along a particle's trajectory and is further weighted by the gamma-ray transmission probability, as discussed in Section~\ref{subsec:gamma_ray_emission_highE}.)

We consider a proton GCR injecting into the flux tube at $\theta_0 = 171^\circ$ because this is where $\langle\mathcal{S}_p\rangle_{\rm tb}$ peaks (see Figure~\ref{fig:P_abs_reflected_GCR}(a)). As for injecting $E_p^{\rm k}$, we choose $15.4$~GeV because proton GCR at this kinetic energy should produce most of the gamma-ray flux at $E_\gamma\sim 1$~GeV. (Note: gamma rays of energy $E_\gamma$ are roughly produced by proton GCRs of kinetic energy $E_p^{\rm k} \sim 10\times E_\gamma$.) In addition, Figure~\ref{fig:P_abs_reflected_GCR}(a) has also shown that $E_p^{\rm k} = 1.54$~GeV and $E_p^{\rm k} = 15.4$~GeV has nearly the same $\langle\mathcal{S}_p\rangle_{\rm tb}$. Thus, the choice of $E_p^{\rm k} = 15.4$~GeV is sufficient for our purpose. Last, when $E_p^{\rm k}$ is as low as $15.4$~GeV, the selections of $\phi_0$ and $r_0$ have minimal influence on $\langle\mathcal{S}_p\rangle_{\rm tb}$ as they yield similar particle trajectories. Thus, we opt for $\phi_0 = 180^\circ$, and $r_0 = 20$~km. Our chosen set of parameters $\left(r_0, \theta_0, \phi_0, E_p^{\rm k}\right)$ for proton GCR injection into flux tube should aptly represent the peak $\mathcal{S}_p$. Therefore, the changes of $\mathcal{S}_p$ observed for this particular set of injection parameters can be regarded as an upper limit on the true impact on the solar gamma-ray emission.

\begin{figure}[t] 
   \centering
   \includegraphics[width=0.98\columnwidth]{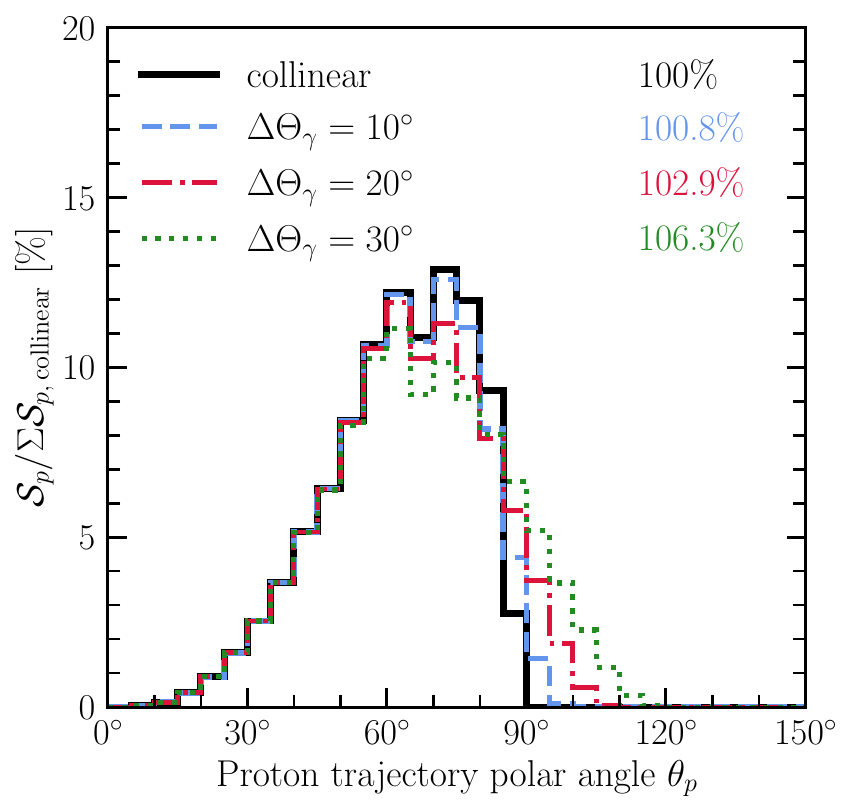}
   \caption{Proton GCR absorption probability in the flux tube weighted by gamma-ray transmission probability under non-collinear gamma-ray emission. The injection polar angle $\theta_0$ is $171^\circ$ and proton GCR $E_p^{\rm k}$ is $15.4$~GeV.}
   \label{fig:S_p_one_particle}
\end{figure}

Figure~\ref{fig:S_p_one_particle} presents our results for $\mathcal{S}_p$ with different angular sizes of emission cones. The $x$~axis is the polar angle of the primary proton GCR along its trajectory. The $y$~axis is $\mathcal{S}_p$ as a function of bins of $\theta_p$ and is further normalized by $\Sigma \mathcal{S}_{p, \, {\rm collinear}}$, which is the summation of $\mathcal{S}_p$ over all bins of $\theta_p$ in the case of collinearity. The black line represents $\mathcal{S}_p/\Sigma \mathcal{S}_{p, \, {\rm collinear}}$ in the case of collinearity. Therefore, the area under the black line equates to $100\%$, as displayed in the top-right corner of the diagram. The black line shows that the gamma rays successfully transmitted from the Sun are produced at $\theta_p< 90^\circ$, echoing the findings in the average angular emission plot in Figure~\ref{fig:mean_emission_angle}.

In Figure~\ref{fig:S_p_one_particle}, color lines show emission cones with $\Delta\Theta_\gamma$ of $10^\circ$, $20^\circ$, and $30^\circ$. All three color lines indicate an increase in $\mathcal{S}_p$ when $\theta_p > \theta_{p,{\rm crit}} \approx 93^\circ$. This increase is due to a fraction of gamma rays leaving the Sun before $\theta_p$ reaches $\theta_{p,{\rm crit}}$. In contrast, when $\theta_p < \theta_{p,{\rm crit}}$, we see a drop in $\mathcal{S}_p$ compared to the black line. This decrease is due to a fraction of gamma rays remaining fully absorbed by the solar gas at this stage.

In Figure~\ref{fig:S_p_one_particle}, the three colored percentages ($100.8\%$, $102.9\%$, and $106.3\%$) correspond to the sum of $\mathcal{S}_p$ across all $\theta_p$ bins, normalized by $\Sigma \mathcal{S}_{p, \, {\rm collinear}}$, for $\Delta\Theta_\gamma$ values of $10^\circ$, $20^\circ$, and $30^\circ$ respectively. These percentages are over $100\%$, which implies that considering a finite-sized emission cone leads to a slightly higher total gamma-ray flux from the solar disk than the collinearity scenario. However, this $\mathcal{S}_p$ increase is about a few percent and should only increase the solar-disk gamma-ray flux by about the same order of magnitude.

\begin{figure}[t] 
   \centering
   \includegraphics[width=0.98\columnwidth]{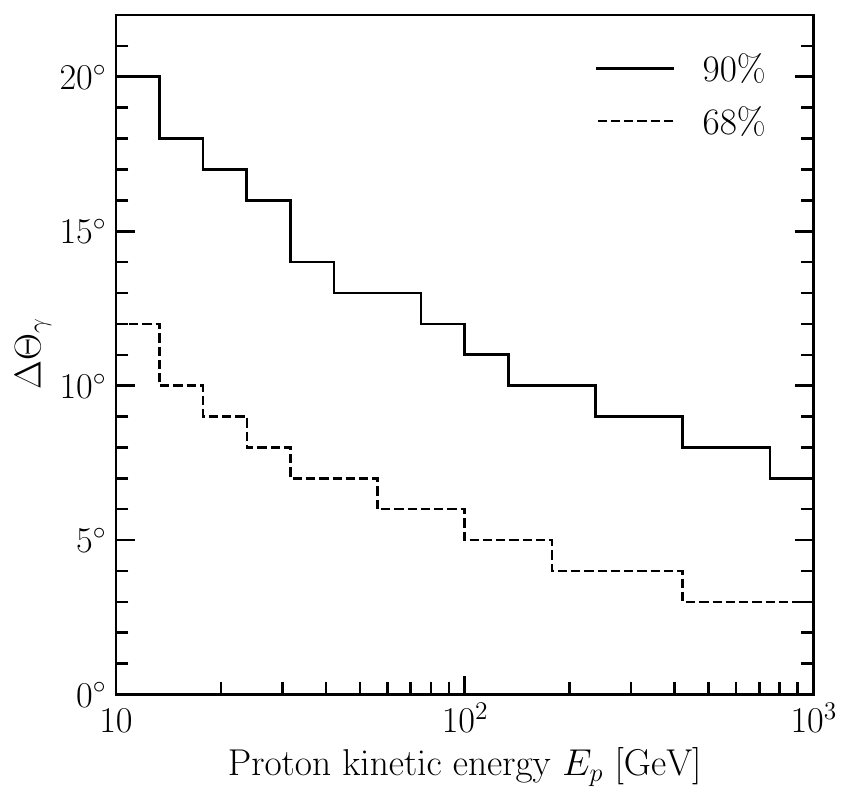}
   \caption{Required cone angles, $\Delta\Theta_\gamma$, to contain $90\%$ and $68\%$ of the total number of gamma rays with $E_\gamma \geq 1$~GeV.}
   \label{fig:FLUKA}
\end{figure}

Last, to estimate the range of gamma-ray emission angles resulting from $pp$ interactions, we utilize the particle shower Monte Carlo simulation code \texttt{FLUKA} (version $\text{4-2.2}$) \citep{2015_Battistoni_fluka, 2022FrP.....9..705A} and its graphical user interface \texttt{Flair} (version $\text{3.2-4.5}$) \citep{Vlachoudis:2009qga}. We inject a proton into a slab consisting of equal numbers of protons and electrons. We calculate the required $\Delta\Theta_\gamma$ that contains either $90\%$ (solid line) or $68\%$ (dashed line) of the total number of gamma rays with $E_\gamma \geq 1$~GeV.

Figure~\ref{fig:FLUKA} shows the result from our \texttt{FLUKA} simulation. The result suggests that $90\%$ of the gamma rays are contained within $\Delta\Theta_\gamma = 20^\circ$ for $E_p^{\rm k} = 10$~GeV, roughly equivalent to peak gamma-ray production for $E_\gamma \sim 1$~GeV. Based on these findings and the $\mathcal{S}_p$ result of $\Delta\Theta_\gamma = 20^\circ$ presented in Figure~\ref{fig:S_p_one_particle}, we conclude that the collinearity assumption should not cause the actual flux to deviate by more than $\approx 3\%$.

\clearpage
\newpage
\bibliography{reference}
\bibliographystyle{aasjournal}

\end{CJK*}
\end{document}